\documentclass{article}

    \PassOptionsToPackage{numbers, compress}{natbib}

\usepackage[eandd, preprint]{neurips_2026}
\usepackage[utf8]{inputenc} 
\usepackage[T1]{fontenc}    
\usepackage{hyperref}       
\usepackage{url}            
\usepackage{booktabs}       
\usepackage{amsfonts}       
\usepackage{nicefrac}       
\usepackage{microtype}      
\usepackage{xcolor}         
\usepackage{tabularx}
\usepackage{array} 
\usepackage{graphicx}
\usepackage{hyperref}
\usepackage{url}
\usepackage{subcaption}
\usepackage{multicol}
\usepackage{multirow}
\usepackage{longtable}
\usepackage{makecell}
\title{BEACON: A Multimodal Dataset for Learning Behavioral Fingerprints from Gameplay Data}

\author{%
  Ishpuneet Singh\thanks{Corresponding author} \\
  Computer Science and Engineering Department\\
  Thapar Institute of Engineering and Technology\\
  Patiala, India \\
  \texttt{isingh\_be22@thapar.edu} \\
  \And
  Gursmeep Kaur \\
  Computer Science and Engineering Department\\
  Thapar Institute of Engineering and Technology\\
  Patiala, India \\
  \texttt{gkaur6\_be22@thapar.edu} \\
  \And
  Uday Pratap Singh Atwal \\
  Computer Science and Engineering Department\\
  Thapar Institute of Engineering and Technology\\
  Patiala, India \\
  \texttt{uatwal\_be22@thapar.edu} \\
  \AND
  Guramrit Singh \\
  Computer Science and Engineering Department\\
  Thapar Institute of Engineering and Technology\\
  Patiala, India \\
  \texttt{gsingh1\_be22@thapar.edu} \\
  \And
  Gurjot Singh \\
  David R Cheriton School of Computer Science\\
  University of Waterloo, Canada \\
  \texttt{g86singh@uwaterloo.ca} \\
  \And
  Maninder Singh \\
  Computer Science and Engineering Department\\
  Thapar Institute of Engineering and Technology\\
  Patiala, India \\
  \texttt{msingh@thapar.edu} \\
}

\begin{document}

\maketitle

\begin{abstract}
Continuous authentication in high-stakes digital environments requires datasets with fine-grained  behavioral signals under realistic cognitive and motor demands. But current benchmarks are often limited by small scale, unimodal sensing or lack of synchronised
environmental context. To address this gap, this paper introduces BEACON (Behavioral Engine for Authentication \& Continuous Monitoring), a large-scale multimodal dataset that captures diverse skill tiers in competitive \textit{Valorant} gameplay. BEACON contains approximately 430 GB of synchronised modality data (461 GB total on-disk including auxiliary \textit{Valorant} configuration captures) from 79 sessions across 28 distinct players, estimated at 102.51 hours of active gameplay, including high-frequency mouse dynamics, keystroke events, network packet captures, screen recordings, hardware metadata, and in-game configuration context. BEACON leverages the high precision motor skills and high cognitive load that are inherent to tactical shooters, making it a rigorous stress test for the robustness of behavioral biometrics. The dataset allows for the study of continuous authentication, behavioral profiling, user drift and multimodal representation learning in a high-fidelity esports setting. The authors release the dataset and code on Hugging Face and GitHub to create a reproducible benchmark for evaluating next-generation behavioral fingerprinting and security models.
\end{abstract}

\section{Introduction and Related Work}

The rapid expansion of interactive digital platforms, encompassing competitive esports, real-money gaming, and persistent online environments, has fundamentally altered the threat surface of human-computer interaction. The increasing time users spend in these high-engagement computing environments makes strong, seamless, and non-intrusive security solutions essential. Traditional point-of-entry authentication mechanisms, such as passwords and one-time PINs, are structurally ineffective for these settings. They rely on one-time verification, so accounts remain vulnerable to session hijacking, credential injection, and account takeover after login. Moreover, standard multi-factor authentication (MFA) prompts break the seamless interaction required in latency-sensitive gaming applications, and the fast-paced nature of competitive play \cite{Nair2023}.
To address this fundamental limitation, continuous authentication has emerged as a frontline
paradigm in cybersecurity \cite{Awasthi2024,  Abbas2026}. This approach operates
silently in the background, persistently validating user identity by analysing behavioral
biometrics such as typing rhythms, touchscreen swipes, and mouse dynamics. Because continuous
authentication relies on subconscious sensorimotor habits rather than memorised secrets, it
offers a ``passwordless'' experience that maintains high security without interrupting user
engagement.
Gaming environments, specifically fast-paced esports like first-person shooters (FPS) and
real-time strategy games, serve as the ultimate crucible for behavioural biometrics. The sheer
frequency, complexity, and intensity of user interactions in these games far exceed those of
traditional web browsing \cite{Awasthi2024, ChenHong2007}. Unlike desktop applications, where user behaviour
is sparse and episodic, competitive games demand relentless, microsecond-level sensorimotor
inputs. These actions, from split-second crosshair flick-shots to complex tactical keybinds,
are deeply tied to an individual's unique cognitive processing, reaction time, and physical
dexterity \cite{Sifa2018}. Previous studies have demonstrated the efficacy of using isolated
modalities, such as mouse dynamics \cite{Balabit2016} and touchscreen gestures
\cite{Sitova2015}, to authenticate users continuously. Similarly, broader research in esports
analytics has successfully employed machine learning techniques to detect cheating, predict
match outcomes, and evaluate player performance based on in-game telemetry
\cite{McIlroyYoung2022, XenopoulosSilva2022, Zimmer2025}.
Despite these promising proofs-of-concept, the scientific community faces a severe bottleneck:
the lack of a comprehensive, multi-modal dataset that captures the full spectrum of behavioral
and environmental telemetry necessary to train and evaluate next-generation foundational AI
models. Existing datasets are often fundamentally constrained. They frequently focus on a single
modality (e.g., only keystrokes \cite{KillourhyMaxion2009}), feature artificially brief
session durations that fail to capture the onset of player fatigue, or originate from
low-stakes environments like free-text typing. Crucially, there is a distinct gap in publicly
available data that precisely correlates raw, event-level hardware inputs with actual
network-level telemetry (e.g., PCAPs) and contextual visual data (e.g., screen recordings) in
high-stakes environments.
To map the current landscape and benchmark the proposed BEACON dataset against existing
literature, Table~\ref{tab:datasets} summarizes prominent behavioral and dynamics datasets
across various domains. The literature reveals a historical reliance on either low-frequency
desktop tasks, unimodal mobile captures, or massive but heavily aggregated enterprise logs.
While recent pioneer efforts like the AMuCS dataset \cite{Fanourakis2025} and mobile-centric
touch databases \cite{nascimento2024} have introduced multimodal affective data for FPS games
and continuous mobile authentication, respectively, BEACON uniquely isolates the critical 
intersection of raw sensorimotor dynamics, hardware configurations, and network telemetry at a large scale.

\begin{table}[!ht]
\centering
\caption{Unified Behavioral \& Dynamics Datasets for Authentication and Profiling}
\label{tab:datasets}
\resizebox{\textwidth}{!}{%
\begin{tabular}{@{}p{4cm}lclllll@{}}
\toprule
\textbf{Dataset Name} & \textbf{Modalities} & \textbf{Subjects} & \textbf{Context / Domain} & \textbf{Task Type} & \textbf{Granularity} & \textbf{Format} & \textbf{Availability} \\ \midrule

\multicolumn{8}{l}{\textbf{\textit{\textbullet~ GAMEPLAY / ESPORTS / VR}}} \\ \midrule
AMuCS Dataset \cite{Fanourakis2025}       & Mouse, Key, Eye, Bio  & 256       & CS:GO (FPS)          & Auth, Profiling   & Event-level   & CSV / EDF     & Public \\
Red Eclipse FPS \cite{RedEclipse2015}     & Mouse, Keyboard       & 10        & FPS Gameplay         & Profiling         & Event-level   & CSV           & Public \\
Dual Intensity \cite{Awasthi2024}         & Mouse                 & 43        & TF2 + Poly Bridge    & Auth              & Event-level   & CSV           & Public \\
Half-Life: Alyx VR \cite{Schell2023}      & Motion, Eye, Bio      & 71        & VR FPS               & Profiling         & Event-level   & CSV           & Public \\
Minecraft Mouse \cite{Siddiqui2021}       & Mouse                 & 40        & Sandbox Gameplay     & Auth              & Event-level   & CSV           & Public \\
Biometric Engagement \cite{Vazquez2022}   & Mouse, Key, Bio       & 62        & Game Engagement      & Profiling         & Aggregated    & CSV           & Public \\
PureSkill.gg \cite{PureSkill2021}         & Game Telemetry        & Millions  & CS:GO Competitive    & Profiling         & Session-level & JSON / SQL    & Public \\
StarCraft II Replay \cite{StarCraft2Dataset} & Game Telemetry     & Millions  & RTS Gameplay         & Profiling         & Session-level & SC2Replay     & Public \\ \midrule

\multicolumn{8}{l}{\textbf{\textit{\textbullet~ DESKTOP (Mouse + Keyboard)}}} \\ \midrule
ISOT Combined \cite{traore2012}      & Mouse, Keyboard       & 24        & Web Browsing         & Continuous Auth   & Event-level   & TXT / CSV     & Public \\
UEBA Webchat \cite{UEBAWebchat2020}       & Mouse, Keyboard       & 11        & Webchat              & UEBA              & Event-level   & CSV           & Public \\
Clarkson II \cite{ClarksonII2017}         & Mouse, Keyboard       & 103       & PC Usage + Gaming    & Continuous Auth   & Event-level   & CSV           & Public \\
KMT Dataset \cite{nnamoko2022}        & Mouse, Key, Touch     & 88        & E-commerce           & Auth              & Event-level   & CSV           & Public \\ \midrule

\multicolumn{8}{l}{\textbf{\textit{\textbullet~ MOUSE DYNAMICS}}} \\ \midrule
Balabit Mouse \cite{Balabit2016}          & Mouse                 & 10        & PC Usage             & Auth              & Event-level   & CSV           & Public \\
SapiMouse \cite{SapiMouse2021}            & Mouse                 & 120       & PC Usage             & Auth              & Event-level   & CSV           & Public \\ \midrule

\multicolumn{8}{l}{\textbf{\textit{\textbullet~ KEYSTROKE DYNAMICS}}} \\ \midrule
CMU Keystroke \cite{KillourhyMaxion2009}  & Keyboard              & 51        & Password Entry       & Auth              & Aggregated    & TXT / CSV     & Public \\
KeyRecs \cite{KeyRecs2023}                & Keyboard              & 99        & Free-text + Password & Auth              & Event-level   & CSV           & Public \\
Liveness Detection \cite{Gonzalez2022}    & Keyboard              & N/A       & Attack Simulation    & Auth              & Mixed         & CSV           & Public \\ \midrule

\multicolumn{8}{l}{\textbf{\textit{\textbullet~ MOBILE / TOUCH}}} \\ \midrule
BrainRun \cite{papamichail2019}              & Touch, Sensors        & $\sim$2000 & Mobile Gameplay     & Auth              & Event-level   & CSV           & Public \\
HMOG \cite{Sitova2015}                    & Touch, Motion, Key    & 100       & Mobile Auth          & Continuous Auth   & Event-level   & CSV           & Public \\
Touch Dynamics \cite{Tehetal2016, Tehetal2020}   & Touch                 & 41        & Mobile Auth          & Auth              & Event-level   & CSV           & Public \\
AuthenTech \cite{nascimento2024}          & Touch                 & 15        & Minecraft (Mobile)   & Auth              & Event-level   & CSV           & Public \\ \midrule

\multicolumn{8}{l}{\textbf{\textit{\textbullet~ PROPOSED DATASET}}} \\ \midrule
\textbf{BEACON} \cite{beacon_dataset} & \textbf{Mouse, Key, Net, Game} & \textbf{28} & \textbf{Valorant (FPS)} \cite{riot_valorant_2020} & \textbf{Auth + Profiling} & \textbf{Event \& Packet} & \textbf{CSV/PCAP/MP4} & \textbf{Released*} \\ \bottomrule
\end{tabular}%
}

{\raggedright \tiny 
\textbf{*}The BEACON dataset is available at: \url{https://huggingface.co/datasets/beacon-gui/BEACON-Dataset} \\
\textbf{Granularity Definitions:} \\
\textbf{Packet-level:} Raw network/video frames. \\
\textbf{Event-level:} Raw timestamped X/Y, clicks, or key timings. \\
\textbf{Action-level:} Grouped gestures/intents (e.g., flick-shots). \\
\textbf{Session-level:} Aggregated statistics (e.g., APM, match telemetry). \par}
\end{table}

To decisively bridge the gap identified in the literature, this paper introduces BEACON
(Behavioral Engine for Authentication \& Continuous Monitoring). Envisioned as a foundational
asset for cybersecurity, BEACON is a large-scale, multi-modal dataset comprising approximately
430~GB of synchronized modality data (461~GB total on-disk including auxiliary \textit{Valorant} configuration captures) collected across 79 real-world sessions and 28 distinct
players. Utilising a custom-built, low-latency logging architecture deployed during live
competitive gameplay, BEACON safely encapsulates over 90 million mouse events, approximately
498,000 keystrokes, and over 114 million network packets. By formally releasing this dataset on
Hugging Face, we aim to equip the machine learning and cybersecurity communities with an
unprecedented, rigorous benchmark for stress-testing continuous authentication systems and
autonomous AI agents under realistic, high-fatigue conditions.
All participants provided informed written consent prior to data collection. The study was
conducted in accordance with the ethical guidelines of \textbf{Thapar Institute of Engineering and Technology},
in compliance with institutional policies for human-subjects research. Participants span a
diverse range of competitive skill tiers on the Valorant ladder, ensuring behavioral diversity
across novice and expert motor profiles. All released data is fully anonymized; participant
identifiers (P001--P028) are pseudonyms with no personally identifiable information retained
in the public dataset.
\section{The BEACON Architecture and Data Collection}
To collect the detailed data needed for behavioral fingerprinting, we developed a custom framework that records player actions without interfering with gameplay. We chose \textit{Valorant} \cite{riot_valorant_2020} as our data source because it is a fast-paced, competitive game that forces players to make split-second decisions under high pressure.
Data collection followed a hybrid approach: most participants were recorded in a controlled lab setting to keep hardware consistent, while others contributed from home to ensure the dataset includes real-world variety. Unlike normal computer work, playing \textit{Valorant} creates a constant stream of rapid mouse movements and keyboard clicks. These actions reveal distinctive behavioral signatures, including reaction time, hand-eye coordination, motor precision, and cognitive load patterns, making it an ideal environment for building secure, continuous authentication systems.

\subsection{Custom Logger Architecture}
\label{sec:logger_arch}

\begin{figure}[!ht]
  \centering
  \scalebox{0.145}{
    \includegraphics[trim={0 0 0 0}, clip]{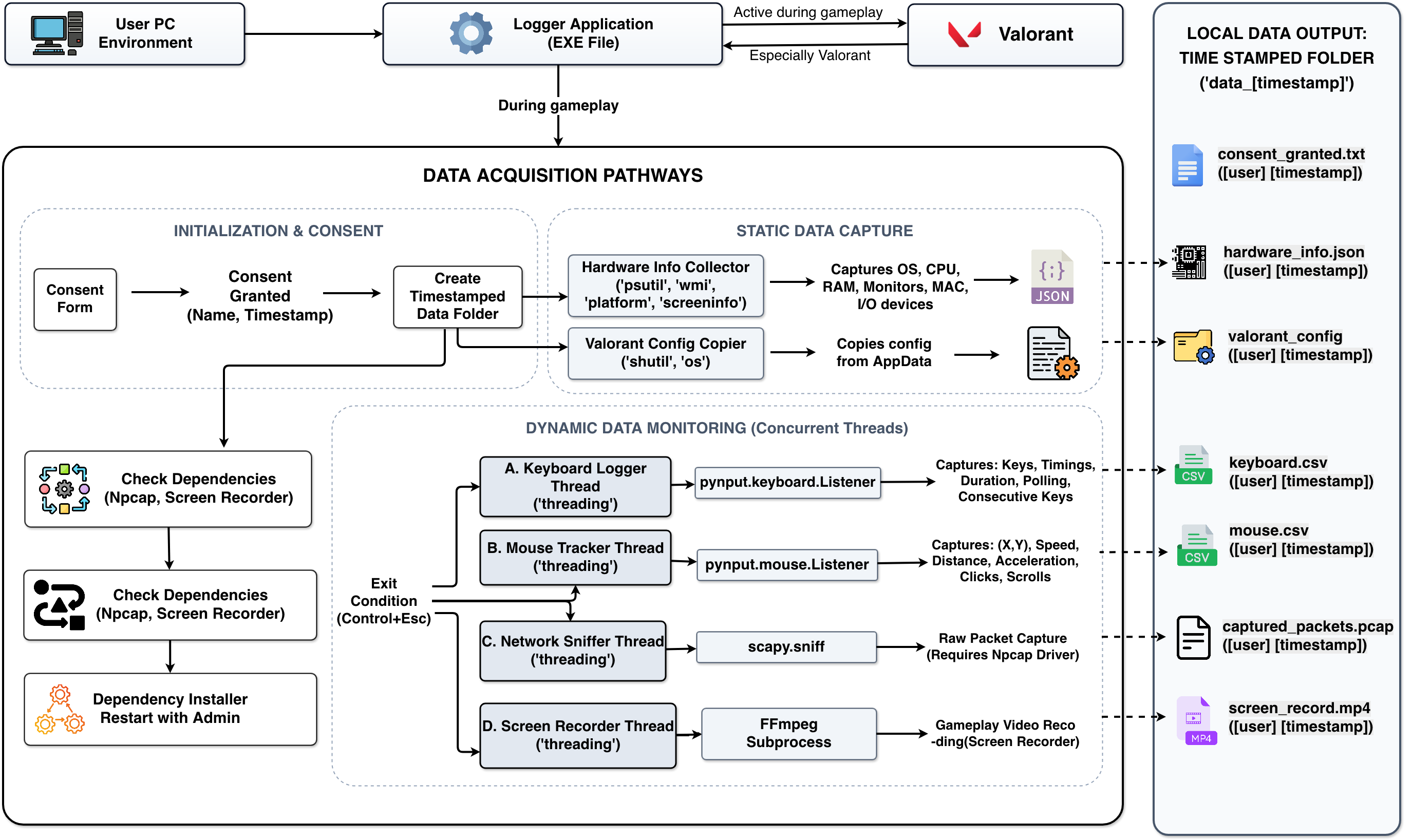}
  }
    \caption{Architecture of the custom BEACON logger \cite{beacon_logger}, detailing the local data acquisition pathways across initialization, static data capture, and concurrent dynamic monitoring threads.}
    \label{fig:logger_arch}
\end{figure}

A core design objective of BEACON was to capture multi-modal gameplay telemetry without degrading frame rate, interrupting the player, or interfering with the normal execution of the game. As shown in Figure~\ref{fig:logger_arch}, the BEACON logger \cite{beacon_logger} was implemented as a standalone executable that runs alongside gameplay and creates a dedicated, time-stamped output directory of the form \texttt{data\_[timestamp]}.

The logging workflow begins with participant consent and session initialization. The logger records a consent artifact (\texttt{consent\_granted\_[timestamp].txt}) and verifies the availability of required dependencies such as packet capture support and screen recording utilities. Following initialization, the logger performs static data capture to preserve the environmental context. A hardware information collector extracts metadata about the CPU, RAM, display configuration, and peripheral environment (\texttt{hardware\_info\_[user]\_[timestamp].json}), while a configuration module copies the local \textit{Valorant} settings to capture customized sensitivities and keybinds.
Once gameplay begins, the logger transitions to dynamic monitoring through four concurrent threads, ensuring multi-modal data acquisition proceeds without blocking the main game loop:
\begin{itemize}
    \item \textbf{Keyboard telemetry:} Captures individual key press/release events, timestamps, dwell times, and inter-key latencies using \texttt{pynput.keyboard.Listener}.
    \item \textbf{Mouse telemetry:} Records cursor coordinates, movement trajectories, click events, scroll activity, speed, and acceleration using \texttt{pynput.mouse.Listener}.
    \item \textbf{Network telemetry:} A \texttt{scapy.sniff} module captures raw network traffic (\texttt{captured\_packets\_[user]\_[timestamp].pcap}) to temporally align local physical inputs with server-side game activity.
    \item \textbf{Screen recording:} An \texttt{FFmpeg} subprocess captures the gameplay session (\texttt{screen\_record\_[user]\_[timestamp].mp4}) using deliberately conservative encoder settings, namely \texttt{libx264} with the \texttt{ultrafast} preset, \texttt{yuv420p} chroma subsampling, and a 25~fps frame rate at native display resolution, to minimise CPU contention with the running game while preserving the visual context required for downstream analysis.
\end{itemize}
All telemetry is written incrementally to disk on a per-event basis throughout the session, ensuring data integrity in the event of an unexpected exit or power loss, resulting in a unified local session package.

\paragraph{Cross-modality temporal alignment.} To support reliable downstream fusion across heterogeneous modalities, every event recorded by the logger is stamped with the host's POSIX time (Unix epoch seconds, via Python \texttt{time.time()}) rather than relative or per-thread offsets. Mouse and keyboard rows, packet capture timestamps emitted by \texttt{scapy}, and the system-clock-driven \texttt{FFmpeg} frame timeline all share this single source of truth, allowing modalities to be aligned post-hoc by intersecting their absolute timestamp ranges without requiring any explicit synchronisation marker.

\subsection{Data Pipeline and Secure Ingestion}
\label{sec:data_pipeline}

Because each gameplay session generates a massive collection of heterogeneous files (often several gigabytes), a dedicated, highly scalable upload architecture was developed to reliably move data from the participant's local device to centralized storage. The BEACON pipeline handles client-side caching, chunked HTTPS transmission to bypass browser limits for large artifacts (e.g., video and PCAP files), and API gateway ingestion. Crucially, a server-side validation layer enforces strict structural checks verifying file byte-sizes, filename regex patterns, and the synchronous presence of all mandatory modalities before migrating the session to finalized database storage for the Hugging Face release. A comprehensive breakdown of the ingestion architecture, secure transport mechanisms, and validation logic, along with detailed schematics, is provided in Appendix \ref{app:data_pipeline}.

\section{Dataset Characteristics and Exploratory Data Analysis}
\label{sec:eda}

The BEACON dataset represents one of the largest publicly available repositories of high-frequency behavioral telemetry. Designed explicitly to facilitate robust machine learning evaluations for continuous authentication, player profiling, and anomaly detection, it captures microsecond-level interactions rather than aggregated session statistics.

\subsection{Overall Statistics and Modality Inventory}

The dataset comprises approximately 430~GB of synchronized modality data (461~GB total on-disk including auxiliary \textit{Valorant} configuration captures) across 79
real-world \textit{Valorant} sessions from 28 distinct participants. The total estimated active
gameplay duration spans approximately 102.51 hours. As detailed in
Table~\ref{tab:modality_inventory}, the raw scale of the captured interactions is vast,
yielding over 90 million distinct sensorimotor events and over 114 million network packets.

\begin{table}[!ht]
\centering
\small
\caption{BEACON Modality Inventory and Storage Distribution}
\label{tab:modality_inventory}
\begin{tabularx}{\textwidth}{@{}Xccc@{}}
\toprule
\textbf{Modality / File Type} & \textbf{Total Files} & \textbf{Data Size} & \textbf{Recorded Events / Rows} \\ \midrule
Screen Recordings (\texttt{screen\_record\_mp4})         & 74 & 381.1 GB  & N/A (Video Context)       \\ \addlinespace[2pt]
Network Telemetry (\texttt{captured\_packets\_pcap})     & 80 & 38.68 GB  & 114,898,144 packets       \\ \addlinespace[2pt]
Mouse Dynamics (\texttt{mouse\_csv})                     & 80 & 9.74 GB   & 90,172,347 events         \\ \addlinespace[2pt]
Keystroke Dynamics (\texttt{keyboard\_csv})              & 80 & 40.57 MB  & 498,801 events            \\ \addlinespace[2pt]
Hardware Context (\texttt{hardware\_info\_json})         & 79 & $<$1 MB   & 79 JSON records           \\ \bottomrule
\end{tabularx}

{\raggedright \footnotesize
\textit{Note:} The 79 collected sessions yield modality file counts that deviate slightly from a strict 1:1 mapping. Screen recordings are absent in 7 sessions (4 early pilot sessions with capture intentionally disabled, 3 due to capture utility failures), and 2 sessions produced split-part recordings, netting 74 video files. One session folder bundles two distinct recording timestamps (March and April 2025) for a single participant, contributing two files per input modality (yielding 80 PCAP, mouse, and keyboard files); the corresponding hardware JSON for the second timestamp was not written, leaving 79 JSON records. \par}
\end{table}

\begin{figure}[!ht]
    \centering
    \scalebox{0.18}{
        \includegraphics[trim={0 0 0 0}, clip]{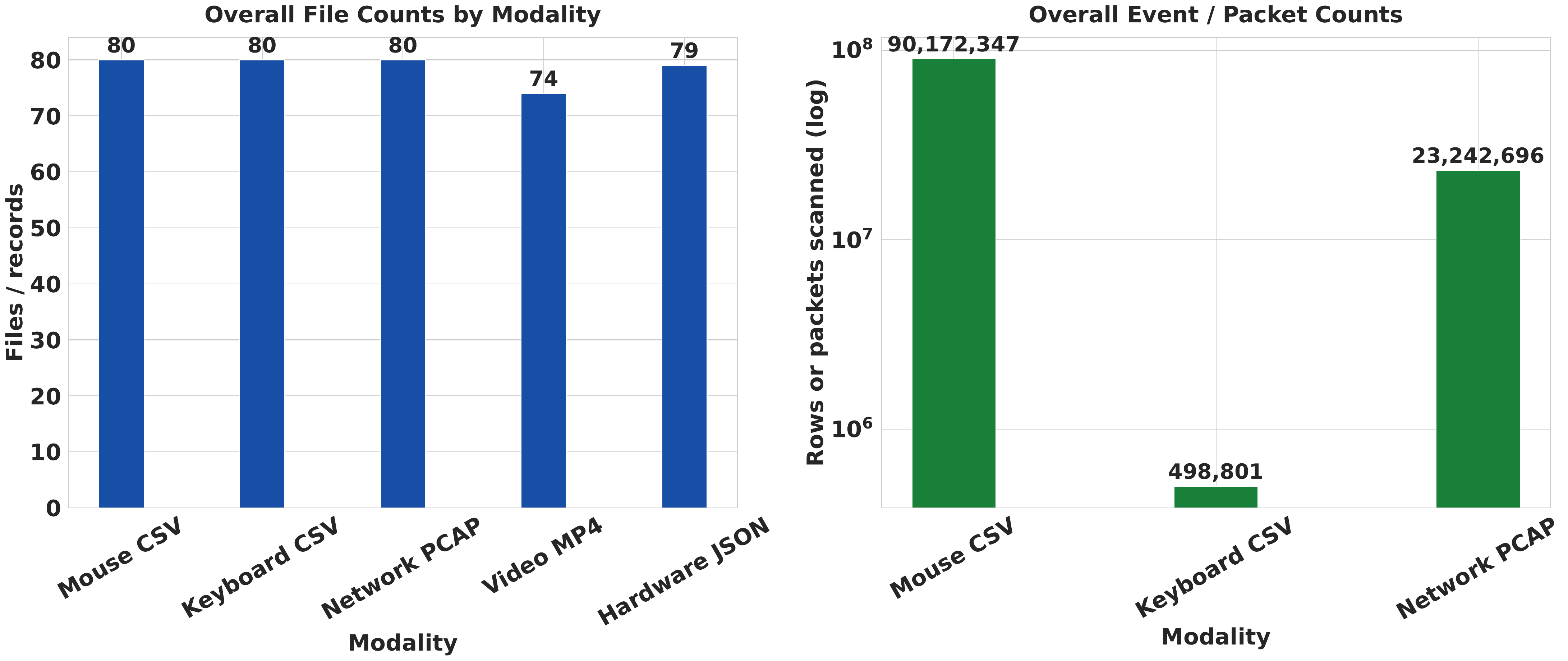}
    }
    \caption{Log-scale distribution of the overall event and packet counts across the primary modalities, emphasising the extreme data density of mouse telemetry relative to keyboard inputs.}
    \label{fig:overall_counts}
\end{figure}

As shown in Figure \ref{fig:overall_counts}, the sheer volume of mouse telemetry dwarfs keystroke dynamics by two orders of magnitude. This reflects the nature of FPS gameplay, where continuous camera movement and rapid crosshair adjustments occur at much higher polling rates than tactical key presses.

\subsection{Exploratory Data Analysis: Spatial and Behavioral Dynamics}
\begin{figure}[!ht]
    \centering
    \begin{subfigure}[b]{1.0\textwidth}
        \centering
        \scalebox{0.14}{
            \includegraphics[trim={0 0 0 0}, clip]{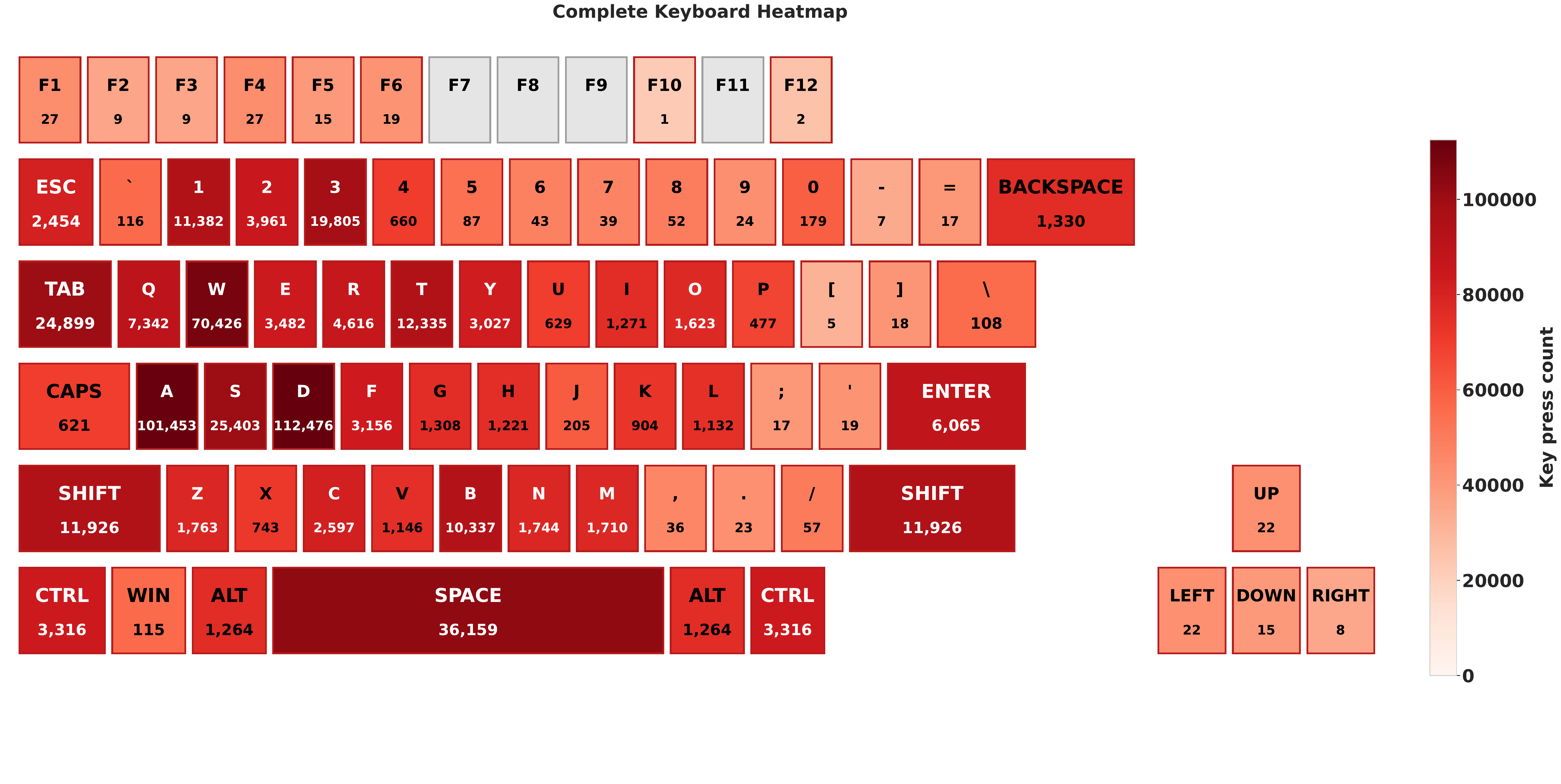}
        }
        \caption{Keyboard Press Frequency}
        \label{fig:heatmaps_a}
    \end{subfigure}
    \par\vspace{1.5em} 
    \begin{subfigure}[b]{1.0\textwidth}
        \centering
        \scalebox{0.32}{
            \includegraphics[trim={1cm 0 0cm 0cm}, clip]{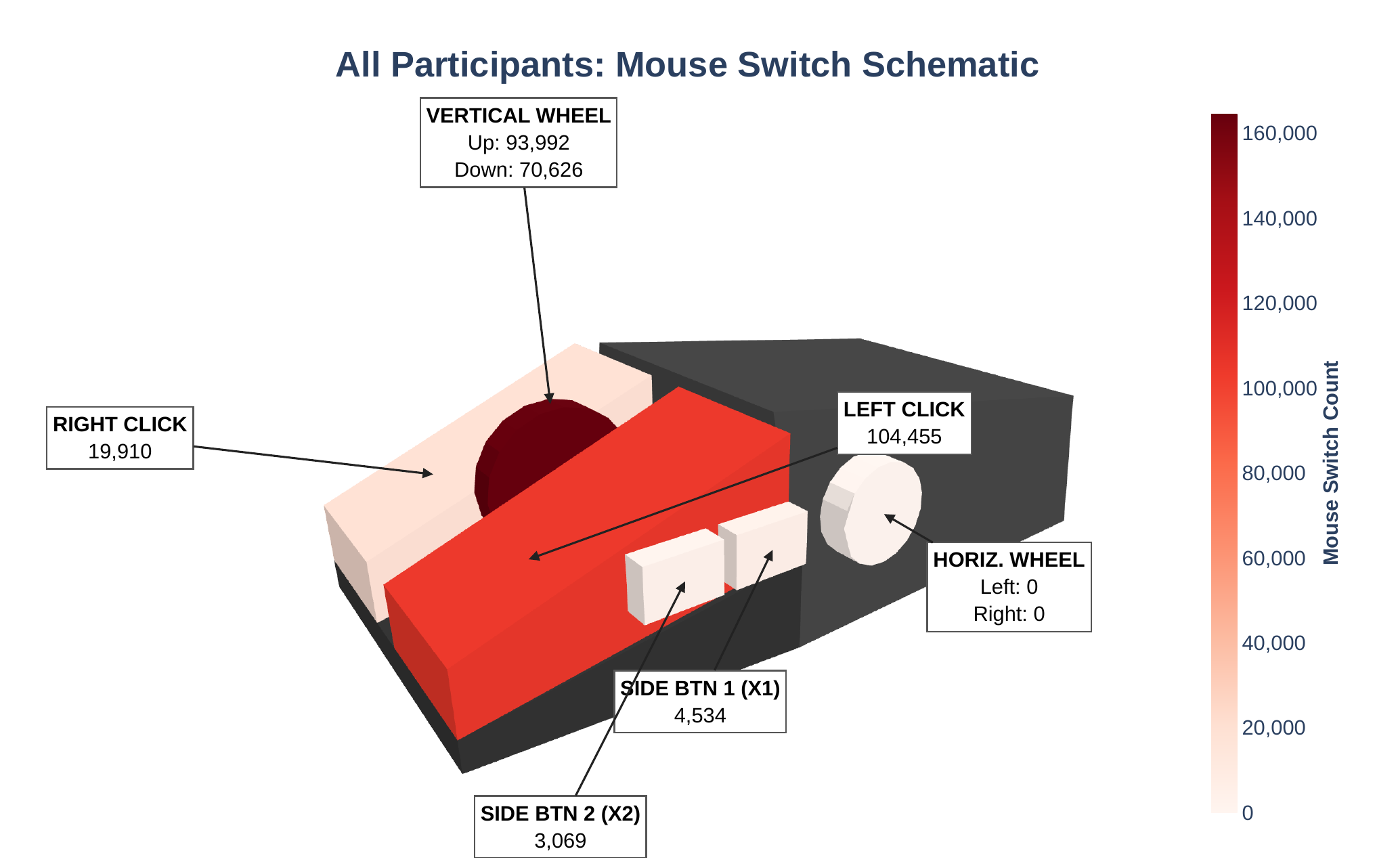}
        }
        \caption{Mouse Switch Usage Frequency}
        \label{fig:heatmaps_b}
    \end{subfigure}
    \caption{Aggregate heatmaps illustrating the spatial distribution of sensorimotor interactions across all 28 participants during active gameplay.}
    \label{fig:heatmaps}
\end{figure}
\begin{figure}[!ht]
    \centering
    \scalebox{0.165}{
        \includegraphics[trim={0cm 0 0 0}, clip]{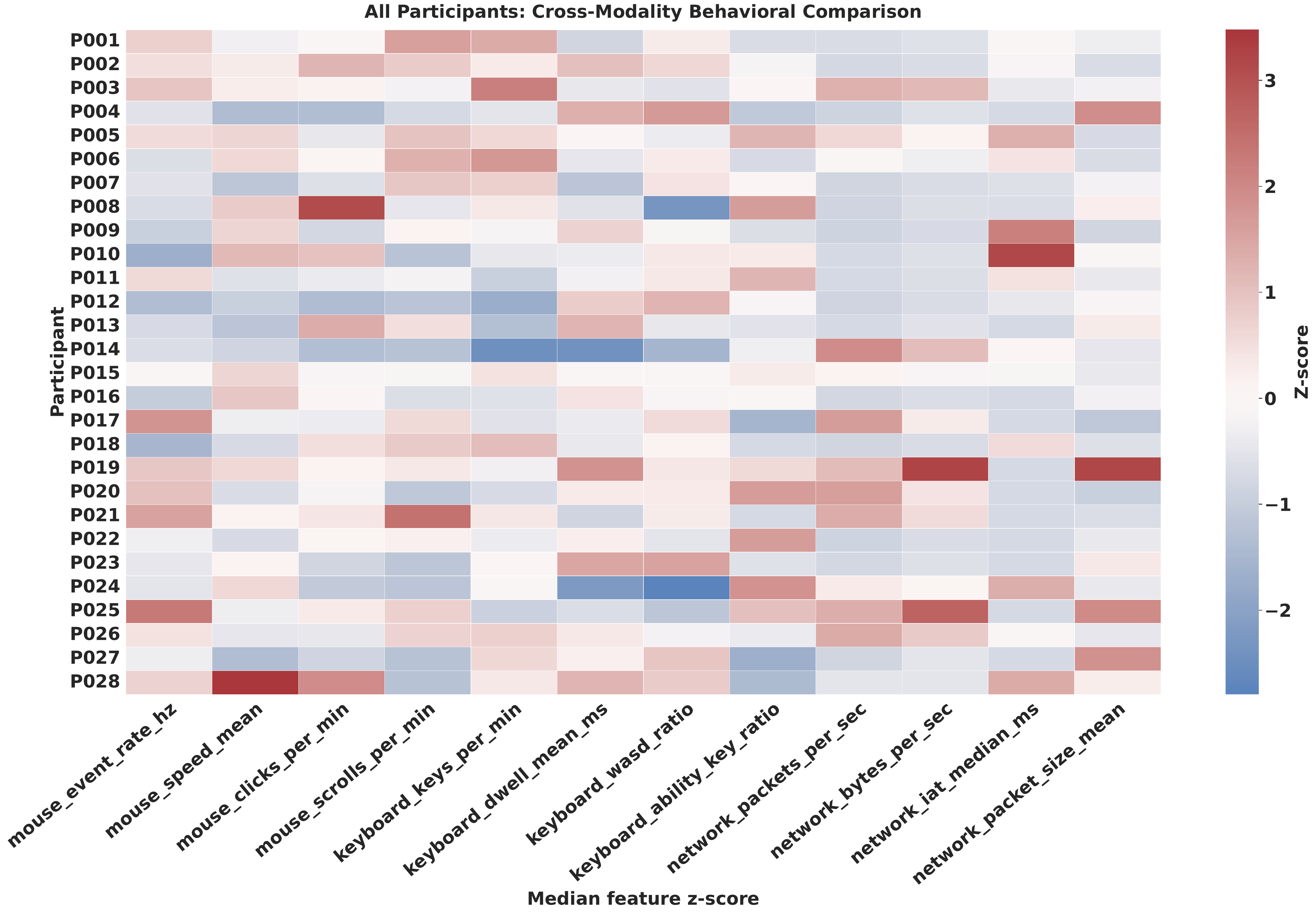}
    }
    \caption{Cross-modality behavioral comparison. The Z-score heatmap standardises median features across all 28 participants, revealing individualised biometric profiles across hardware interactions.}
    \label{fig:zscore_heatmap}
\end{figure}

Initial exploratory data analysis underscores the complexity, variance, and biometric viability of the captured telemetry. Figure \ref{fig:heatmaps} illustrates the spatial frequency of inputs across the hardware. The complete keyboard heatmap (Figure \ref{fig:heatmaps_a}) confirms that user interactions are heavily concentrated around the \texttt{W-A-S-D} movement cluster, alongside tactical binds like \texttt{SHIFT} and \texttt{SPACE}. Similarly, the mouse switch usage heatmap (Figure \ref{fig:heatmaps_b}) reveals that while primary firing actions (Left Mouse Button) naturally dominate the distribution, the utilisation of secondary interactions such as scoping (Right Mouse Button) and specialised scroll-wheel mechanics varies significantly among players depending on their in-game roles and physical habits.
Despite this structural similarity dictated by core game mechanics, participants possess genuinely distinct behavioral signatures across modalities. Figure \ref{fig:zscore_heatmap} standardises core features (e.g., mouse speed, key press rate) into a cross-modality Z-score heatmap. The distinct horizontal banding proves that players maintain highly individualised profiles. For instance, a player might exhibit aggressively fast mouse speeds but surprisingly low keyboard interaction rates, a combination uniquely identifiable to a machine learning classifier.

\subsection{Individual Distinctiveness and Biometric Separability}

\begin{figure}[!ht]
    \centering
    \scalebox{0.17}{
        \includegraphics[trim={0 0 0 0}, clip]{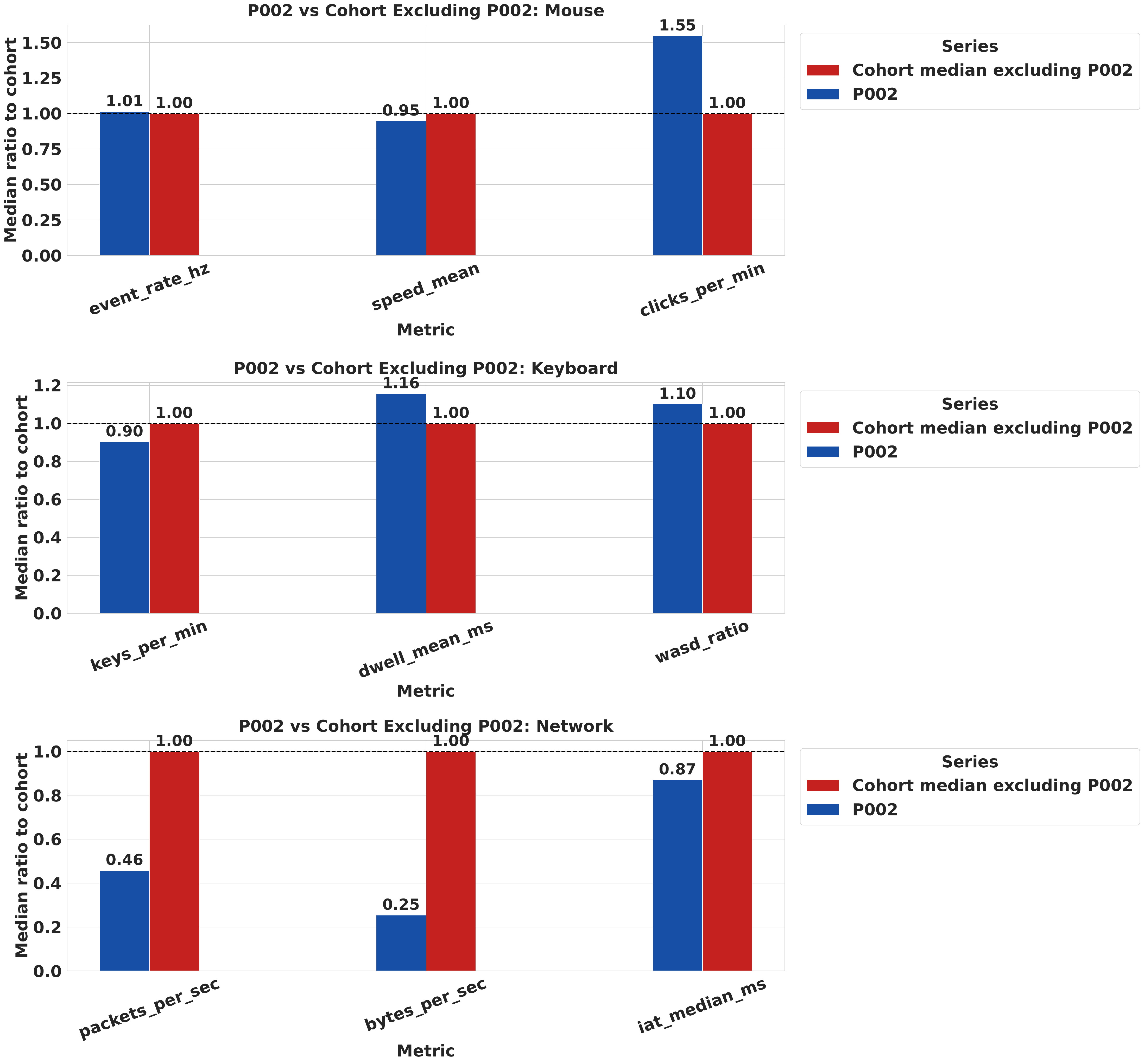}
    }
    \caption{Inter-user separability analysis. The distribution of sensorimotor features for a
    single participant (P002) plotted against the global dataset variance, highlighting tight
    operational boundaries consistent with strong biometric discriminability.}
    \label{fig:p002_vs_all}
\end{figure}

A critical requirement for continuous authentication is that an individual's biometric
signature must be easily distinguishable from the global population (high inter-user variance,
low intra-user variance). As demonstrated in Figure~\ref{fig:p002_vs_all}, comparing a single
user (P002) against the aggregated data of all other players reveals profound separability
across multiple sensorimotor dimensions. While the global distribution contains wide variance
representing the diverse mechanics of the entire player base, P002 maintains a tight, highly
specific operational distribution, a pattern consistently observed across participants
regardless of skill tier or hardware configuration.
This separability is not incidental. It emerges from the deeply habitual nature of
sensorimotor behaviour under competitive cognitive load: each player develops idiosyncratic
aiming mechanics, reaction cadences, and movement rhythms that persist across sessions and
remain stable even under fatigue. The combination of high inter-user variance and low
intra-user variance across both mouse and keyboard modalities confirms that BEACON captures a 
genuine biometric signal rather than session-level noise. The granular statistical
distributions driving this separability across all 28 players are extensively detailed in
Appendix~\ref{app:eda_boxplots}. Ultimately, this multi-dimensional distinctiveness forms the
foundational basis for the baseline evaluation tasks presented in
Section~\ref{sec:results}.

\section{Results}
\label{sec:results}

The baseline performance of the BEACON dataset was computed by transforming raw, asynchronous
telemetry into a structured 28-class identification task. Evaluation focused on six
state-of-the-art architectures originally developed for Website Fingerprinting (WF):
\textbf{ARES \cite{ARES}, BAPM \cite{BAPM}, NetCLR \cite{NetCLR}, TCN \cite{TCN},
TMWF \cite{TMWF}, and Var-CNN \cite{VarCNN}}. These models were selected due to their proven
capacity to model complex temporal dependencies in noisy, high-frequency time-series data.
All architectures were implemented in PyTorch and trained on an NVIDIA H100 80GB GPU.\footnote{We thank the Thapar School of Advanced AI and Data Science for providing the computational resources used in this work.} Input traces were padded or truncated to a fixed sequence length of 1024 tokens. Models were trained for up to 30 epochs with a batch size of 32 using the Adam optimizer (lr = $10^{-3}$) and CrossEntropyLoss. Data was partitioned chronologically: 80\% for training, with 10\% of the training split reserved for validation, and the remaining 20\% held out for final testing. PCAP and screen recording modalities are released as part of BEACON but are deliberately scoped out of the present baselines: this paper's primary contribution is the dataset itself, and the included baselines are intended to characterise mouse and keyboard separability rather than to exhaustively benchmark every modality. Network and video-based identification are explicitly framed as open directions for the community.
Results were computed across three modality configurations: Only Mouse, Only Keyboard, and a
``Combined'' setup featuring both. For each configuration, statistical features were aggregated
across four temporal resolutions: 10s, 30s, 45s, and 60s. A complete dictionary of the 33 engineered features extracted for these baselines is provided in Appendix
\ref{app:feature_dictionary}. The definitions and biometric significance of these metrics are
detailed in Appendix \ref{app:metrics_definitions}. A comprehensive breakdown of all
evaluation metrics, including Accuracy, Equal Error Rate (EER), d-prime ($d'$), and ROC AUC,
is provided in Appendix \ref{app:detailed_results_table}. The corresponding performance and
convergence curves are detailed separately in Appendix \ref{app:evaluation_graphs}.
As summarised in the performance analysis, unimodal mouse dynamics consistently outperformed
unimodal keyboard dynamics. Models trained exclusively on keyboard features peaked at an
identification accuracy of 36.23\% (TMWF, 45s window), while mouse-only models achieved
63.16\% accuracy even at a 10-second resolution (Var-CNN). The early fusion of both modalities
yielded the highest overall performance, with \textbf{Var-CNN} achieving the global maximum
identification accuracy of \textbf{70.82\%} with a \textbf{4.31\% EER} in the 60-second
fusion setup. Conversely, the ARES architecture failed to converge, resulting in 0.00\%
accuracy across all configurations.

\section{Discussion}
\label{sec:discussion}

The experimental results obtained from the BEACON dataset establish several critical paradigms
regarding the nature of high-fidelity behavioral modelling. A primary finding of this study is
that information density serves as the fundamental driver of biometric separability. The
pronounced performance delta observed between mouse and keyboard modalities suggests that
continuous, high-frequency inputs, specifically cursor trajectories, velocity profiles, and
acceleration derivatives, provide a vastly more individualised signature than discrete,
asynchronous tactical key presses. While keyboard interactions are effective at capturing
high-level strategic intent, such as utility usage frequency and movement pacing, they lack
the granularity required for rapid identification. In contrast, the continuous sensorimotor
loop inherent in aiming mechanics appears significantly more resilient to noise and behavioral
mimicry, as evidenced by the consistently higher $d'$ scores in mouse-centric configurations.
Furthermore, the positive scaling observed as temporal windows expand from 10 to 60 seconds
highlights a fundamental ``micro-versus-macro'' trade-off in behavioral biometrics. Shorter
windows effectively isolate micro-reflexes, such as raw mechanical responses during flick-shots
or rapid target re-acquisition. While these features are highly distinctive, they are
susceptible to high intra-user variance induced by immediate in-game stressors. Conversely,
60-second windows encapsulate macro-behaviors, including rotational pacing and habitual
crosshair placement. The convergence of architectures like Var-CNN at these longer durations
suggests that while micro-reflexes provide a rapid identity signal, macro-behaviors provide
the contextual stability necessary to minimise False Rejection Rates in practical,
non-intrusive security deployments.
Finally, the disparate performance among the six adapted website fingerprinting architectures
indicates that specific inductive biases are requisite for modelling human telemetry. The success of Var-CNN and NetCLR implies that dilated causal convolutions and contrastive representation learning is superior for extracting the long-tail temporal dependencies found
in esports data. In contrast, the systematic failure of the ARES framework to converge
highlights a significant domain gap; architectures optimized for the transactional, bursty
timing patterns of encrypted network traffic require substantial structural re-engineering to
accommodate the overlapping, multi-dimensional variance inherent in continuous human
sensorimotor output, where the signal is dense and continuous rather than transactional.

\section{Limitations}
\label{sec:limitations}

While BEACON represents a significant advance in behavioral biometric datasets, several
limitations merit acknowledgment. First, BEACON is positioned as a depth-focused FPS biometrics dataset rather than a population-scale authentication benchmark. The cohort of 28 participants yields approximately 102 user-hours of dense, four-modality telemetry, with per-user event densities (e.g., $\sim$14{,}000 mouse events per minute) that are an order of magnitude higher than typical desktop-biometrics corpora. This design supports per-user behavioral modelling, intra-session stability analysis, and multi-modal fusion research, but it does not support population-level generalisation claims, demographic representativeness, or large-scale enrolment studies; cohorts on the order of $10^2$--$10^3$ participants from broader demographic and skill-tier distributions would be required for those settings. Second, as all sessions were conducted exclusively
on Windows-based gaming PCs running \textit{Valorant}, the cross-platform and cross-game
transferability of learned behavioral fingerprints remains an open question, and analogous
datasets from other FPS titles or operating systems would be necessary to assess
generalization. Third, the present baselines evaluate mouse and keyboard modalities independently and in early fusion; screen recordings and PCAP captures are released as part of the dataset but are intentionally not benchmarked here, as exhaustive multimodal benchmarking lies outside the scope of an initial dataset release and is explicitly framed as an open direction for the community. Finally, this dataset captures signatures at a single point in time;
longitudinal collection spanning months or years would be required to model behavioral drift
as users mature in skill or undergo physical changes over time.

\section{Conclusion}
\label{sec:conclusion}

This study introduced BEACON, a novel, large-scale multimodal dataset designed to advance
behavioral biometrics through high-fidelity esports telemetry. By synchronizing approximately
430~GB of modality telemetry (461~GB total on-disk) across 79 real-world sessions and 28 distinct players, the dataset
provides a rigorous platform for evaluating continuous authentication under extreme cognitive
load. The experimental phase established that deep learning architectures adapted from website
fingerprinting can effectively identify users with high precision even at short temporal
resolutions, with mouse-derived motor signatures consistently emerging as the most
discriminative unimodal feature. The early fusion of hardware modalities further demonstrated
that combining complementary behavioral signals yields the most resilient identification
performance, reducing Equal Error Rates and improving biometric separability across all
evaluated architectures. Ultimately, BEACON establishes a foundational benchmark for the
transition from static, point-of-entry security to dynamic, continuous verification
frameworks, equipping the research community with the data infrastructure necessary to develop
the next generation of non-intrusive, adaptive authentication systems.

\section{Future Scope}
\label{sec:future_scope}

A primary direction is adversarial robustness: characterising the susceptibility of behavioral signatures to mimicry attacks and AI-driven bots that emulate a target's aim-path curvature, and developing liveness-detection mechanisms that distinguish authentic human interactions from synthetic replicas. Baseline performance can be further improved through expanded feature engineering and sliding-window architectures with varying overlap ratios. A natural follow-up is the direct fusion of network-level PCAP data and high-resolution video frames into the feature space, combining computer-vision spatial context with packet-level jitter analysis to yield biometric profiles that are substantially harder to spoof. The data density of BEACON also enables self-supervised behavioral foundation models, with applications beyond security to skill assessment, fatigue detection, and personalised coaching. Finally, broader demographic and hardware coverage and longitudinal collection over months or years are essential to model behavioral drift and ensure long-term reliability of continuous-authentication systems in deployment.

\clearpage
\bibliographystyle{plainnat}
\bibliography{ref.bib}

\newpage
\appendix

\section{Ethics Statement}
\label{app:ethics}

\paragraph{Consent, oversight, and anonymisation.} All participants were adults and provided informed written consent prior to each recording session, including explicit acknowledgment that mouse, keyboard, network, screen, and hardware telemetry would be captured. Data collection was conducted under the ethical guidelines of the host institution for human-subjects research. The publicly released dataset is anonymised through the modality-specific pipeline detailed in \S\ref{subsec:datasheet_distribution}: per-session names are replaced with opaque pseudonymous identifiers (P001--P028) via a private salt-keyed mapping that is held by the curators and never published; hardware JSONs are stripped of hostname, IP, and MAC fields; PCAPs are released with IP/MAC addresses remapped and local-network discovery traffic dropped; and screen recordings are released with audio removed and identifying on-screen regions masked. Raw consent artefacts are retained privately and are not redistributed. The mapping retention policy supports targeted withdrawal: a participant may request removal of all of their sessions without disrupting the remainder of the dataset.

\paragraph{Sensitive content in captures.} Screen recordings can incidentally include on-screen content beyond gameplay (e.g., desktop notifications, taskbar items, browser tabs visible during alt-tabs). Participants were specifically informed of this risk during consent, and sessions were terminated and discarded on request when participants flagged inadvertent capture of personal content. The release-side video pipeline masks these surfaces together with identifying in-game elements. No webcam, microphone, or biometric physiological signal was recorded.

\paragraph{Dual-use considerations.} Behavioral biometrics is an inherently dual-use technology: the same signals that enable continuous authentication also enable cross-context user tracking and surveillance. We release BEACON to enable defensive research, including continuous authentication, anti-cheat, and bot/mimicry detection, but we explicitly discourage uses such as covert behavioral profiling, re-identification across services, or worker-monitoring systems deployed without informed consent. Researchers using BEACON are expected to comply with applicable data-protection regulations in their jurisdiction and to seek their own ethical review for any downstream study involving human subjects.

\paragraph{Game-platform terms.} BEACON contains gameplay artefacts derived from \textit{Valorant}, a proprietary title operated by Riot Games. No proprietary game assets, decrypted protocol data, or anti-cheat internals are included. Researchers should independently verify compliance with Riot Games' terms of service for their specific use case.

\newpage
\section{Extended Data Pipeline and Secure Ingestion Details}
\label{app:data_pipeline}

\begin{figure}[!ht]
  \centering
  \scalebox{0.18}{
    \includegraphics[trim={0 0 0 0}, clip]{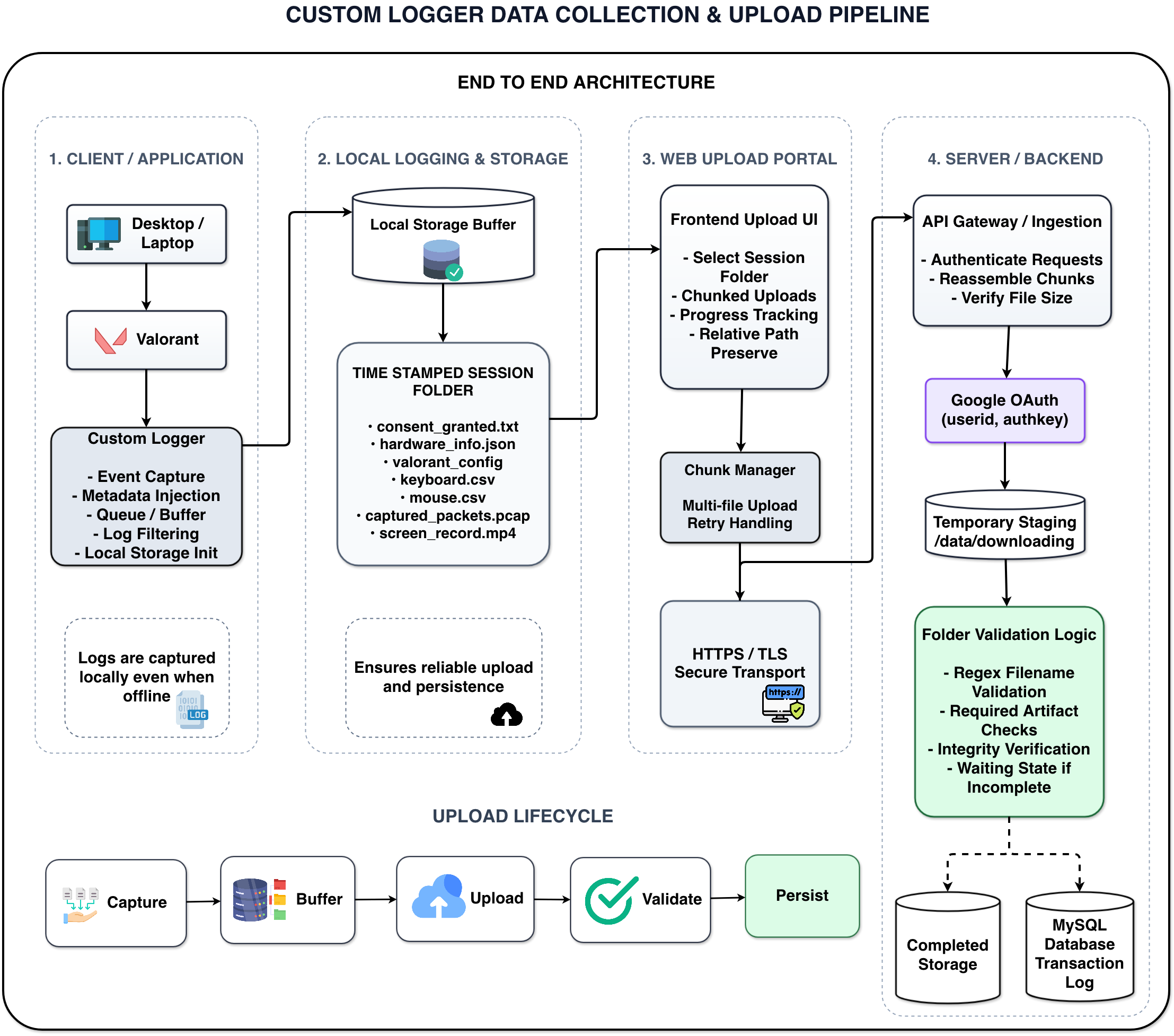}
  }
    \caption{The secure data pipeline and upload portal, illustrating local caching, chunked HTTPS transmission, API gateway ingestion, and server-side validation logic.}
    \label{fig:upload_portal}
\end{figure}

To manage the massive scale of the multimodal data collected by BEACON, a dedicated upload architecture was engineered. As illustrated in Figure~\ref{fig:upload_portal}, the pipeline consists of four stages: client-side collection, local buffering, upload portal processing, and server-side validation and storage.

At the client side, the logger records all artifacts locally before any transfer occurs. Once a session is complete, the participant uses a custom frontend portal to submit the time-stamped session folder. Since browser-based upload workflows are often unreliable for gigabyte-scale files, the system implements a chunked upload protocol for large artifacts (e.g., \texttt{.mp4} and \texttt{.pcap} files) while preserving relative folder paths. 
All uploads are transmitted over an HTTPS/TLS-secured channel to the server backend. Incoming requests are authenticated via OAuth, and the uploaded fragments are temporarily placed into a staging area where they are reassembled server-side. A dedicated validation layer then performs structural checks before final acceptance. The backend verifies strict filename patterns, checks whether all mandatory modalities are present, and flags incomplete uploads for review. Sessions that do not satisfy the expected schema are retained in a temporary waiting state rather than being merged into the finalized corpus. Validated sessions are finally migrated to permanent storage, and session metadata (upload status, file counts, sizes) is committed to a MySQL transaction log to monitor storage health and dataset completeness.

\newpage
\section{Extended EDA: Statistical Distributions of Modalities}
\label{app:eda_boxplots}

To further demonstrate the high inter-user variance necessary for effective continuous authentication, the statistical distributions of distinct sensorimotor and system features across all 28 participants were analyzed. A foundational premise of behavioral biometrics is that while an individual user's actions should remain relatively consistent (low intra-user variance), the differences between any two users must be statistically significant (high inter-user variance). The boxplots generated from the BEACON dataset strongly validate this premise across all captured modalities.

\subsection{Keyboard Dynamics Distribution}
\begin{figure}[!ht]
    \centering
    \scalebox{0.18}{
        \includegraphics[trim={0 0 0 0}, clip]{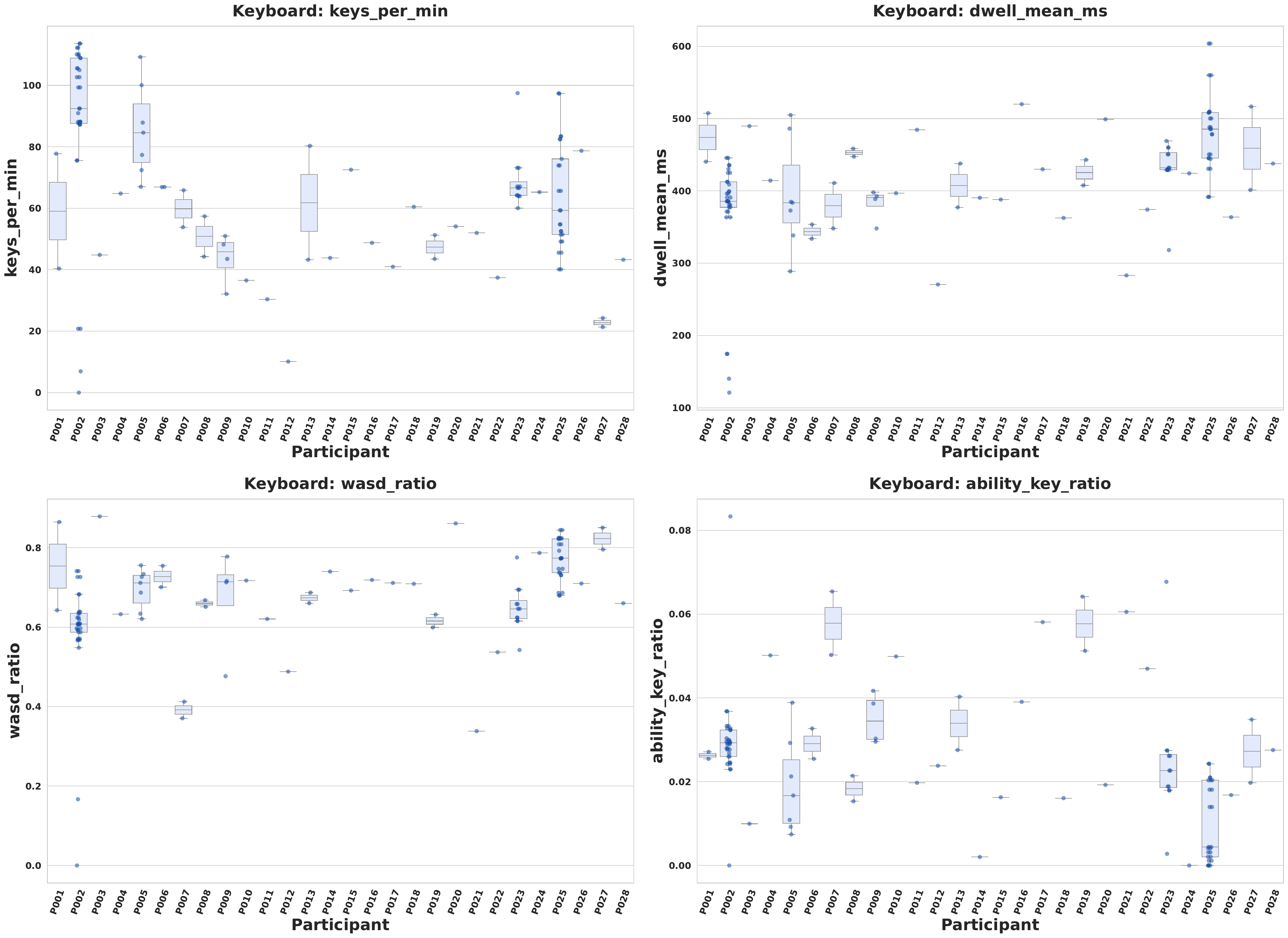}
    }
    \caption{Statistical distribution of Keyboard dynamics across all 28 participants. Notice the high inter-user variance in Mean Dwell Time (ms), which acts as a strong biometric discriminator.}
    \label{fig:app_keyboard_boxplots}
\end{figure}

Figure \ref{fig:app_keyboard_boxplots} illustrates the distribution of four engineered keyboard features: \texttt{keys\_per\_min}, \texttt{dwell\_mean\_ms}, \texttt{wasd\_ratio}, and \texttt{ability\_key\_ratio}. 

\begin{itemize}
    \item \textbf{Physiological Traits (\texttt{dwell\_mean\_ms}):} The mean dwell time the duration a key is held down before release is a classic biometric indicator of physiological dexterity. The boxplots reveal that some participants consistently execute rapid, lightweight taps (medians around 50–100 ms), while others exhibit heavier, prolonged keystrokes (medians exceeding 200 ms). 
    \item \textbf{Tactical Profiling (\texttt{keys\_per\_min} and \texttt{wasd\_ratio}):} Metrics like total keys pressed per minute and the ratio of movement keys (W, A, S, D) to overall keystrokes capture high-level gameplay styles. The tight interquartile ranges (IQRs) for individual players on these metrics suggest that tactical pacing is a deeply ingrained, habitual trait.
\end{itemize}

\subsection{Mouse Telemetry Distribution}
\begin{figure}[!ht]
    \centering
    \scalebox{0.18}{
        \includegraphics[trim={0 0 0 0}, clip]{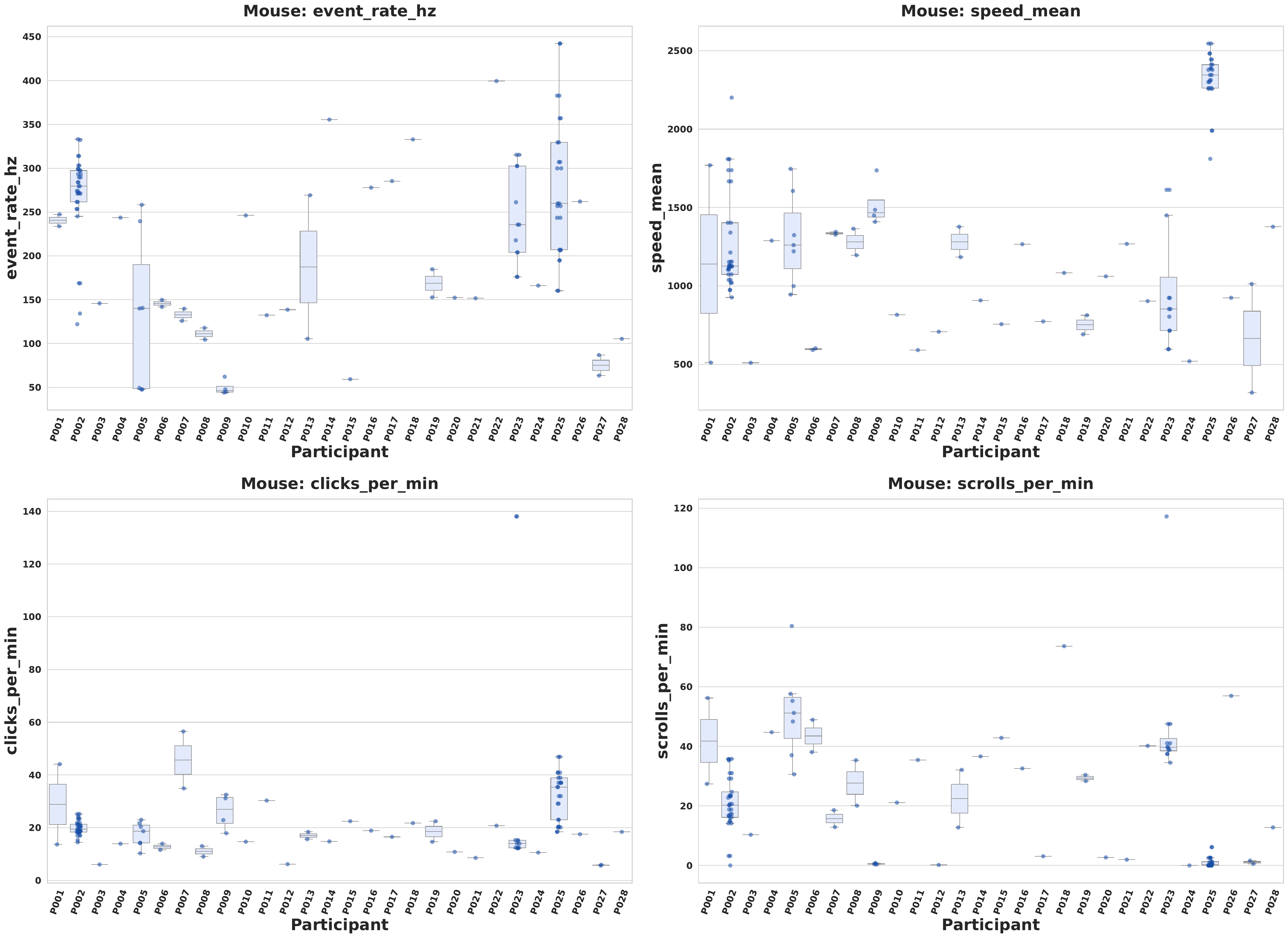}
    }
    \caption{Statistical distribution of Mouse dynamics across all 28 participants. Features such as Event Rate (Hz) and Mean Speed highlight distinct physical hardware capabilities and individual aiming mechanics.}
    \label{fig:app_mouse_boxplots}
\end{figure}

Similarly, Figure \ref{fig:app_mouse_boxplots} presents the distribution of mouse telemetry features: \texttt{event\_rate\_hz}, \texttt{speed\_mean}, \texttt{clicks\_per\_min}, and \texttt{scrolls\_per\_min}.

\begin{itemize}
    \item \textbf{Hardware and Motion Signatures (\texttt{event\_rate\_hz} and \texttt{speed\_mean}):} The \texttt{speed\_mean} directly correlates with a player's in-game sensitivity and physical aiming mechanics (e.g., executing fast, broad sweeps versus slow, precise tracking movements). The severe divergence in median speeds across the 28 participants demonstrates that aiming mechanics are uniquely individualized.
    \item \textbf{Action Frequencies (\texttt{clicks\_per\_min} and \texttt{scrolls\_per\_min}):} Trigger discipline is reflected in the click rates, while scroll patterns often reveal specialized keybinds (e.g., binding jump to the scroll-wheel). The presence of heavy outliers in \texttt{scrolls\_per\_min} for specific participants provides highly separable edge-case features for anomaly detection algorithms.
\end{itemize}

\subsection{Network Telemetry Distribution}
\begin{figure}[!ht]
    \centering
    \scalebox{0.18}{
        \includegraphics[trim={0 0 0 0}, clip]{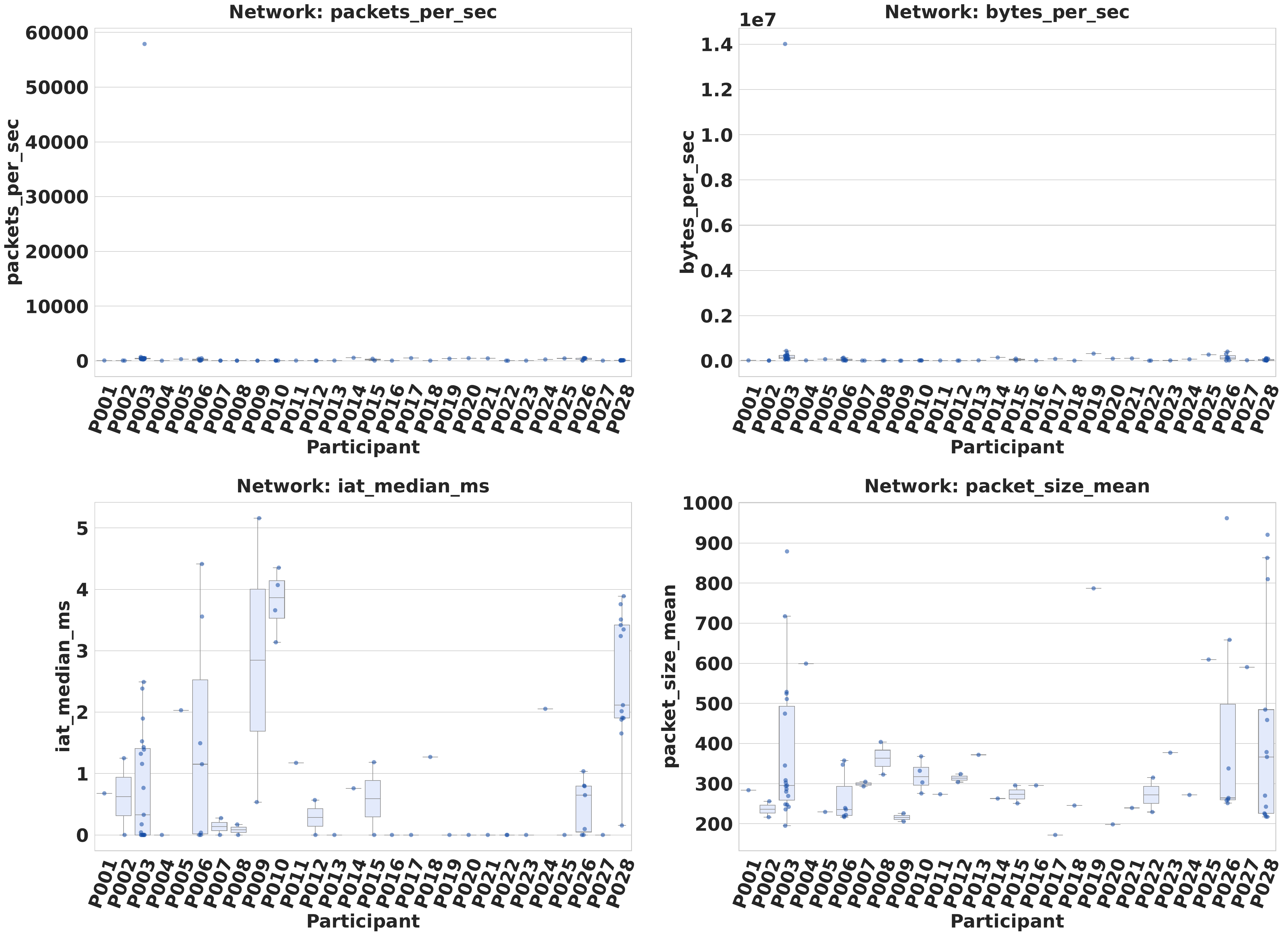}
    }
    \caption{Statistical distribution of Network telemetry across all 28 participants. The integration of packet-level features provides both an environmental baseline and a measure of behavioral throughput.}
    \label{fig:app_network_boxplots}
\end{figure}

Finally, Figure \ref{fig:app_network_boxplots} illustrates the distribution of network-level features extracted from the synchronized PCAP files, serving as the bridge between physical intent and system-level execution.

\begin{itemize}
    \item \textbf{Behavioral Throughput (Packet Rate):} While network traffic is dictated by the game's server tick rate, the volume of outbound packets often correlates with a player's physical action density. Highly aggressive, high-APM (Actions Per Minute) players generate denser bursts of UDP traffic compared to passive players holding strategic angles, creating a secondary layer of behavioral profiling.
    \item \textbf{Environmental Fingerprinting (Inter-Arrival Time):} Features such as packet inter-arrival time and jitter provide an environmental baseline. In the context of continuous authentication, even if an adversarial agent perfectly mimics a user's mouse trajectories, the underlying network variance (dictated by hardware, ISP routing, and geographic location) provides a highly resilient, distinct verification metric that is exceptionally difficult to spoof.
\end{itemize}

\clearpage
\section{Feature Engineering Dictionary}\label{app:feature_dictionary}

To transform the asynchronous raw telemetry into a structured format for machine learning baselines, 33 statistical features were engineered across mouse and keyboard modalities. Table \ref{tab:feature_dictionary} details the naming convention and description for each feature extracted within the temporal windows.

\small
\begin{longtable}{@{}p{0.3\textwidth} p{0.65\textwidth}@{}}
\caption{Summary of Engineered Features extracted for BEACON Baselines.} \label{tab:feature_dictionary} \\
\toprule
\textbf{Feature Name} & \textbf{Description} \\
\midrule
\endfirsthead
\multicolumn{2}{c}{{\bfseries \tablename\ \thetable{} -- continued from previous page}} \\
\toprule
\textbf{Feature Name} & \textbf{Description} \\
\midrule
\endhead
\midrule
\multicolumn{2}{r}{{Continued on next page...}} \\
\endfoot
\bottomrule
\endlastfoot
mouse\_event\_count & Mouse telemetry rows in the window. \\
mouse\_event\_rate & Mouse telemetry rows per second. \\
move\_count & Mouse move events in the window. \\
move\_rate & Mouse move events per second. \\
left\_click\_count & Left-click events in the window. \\
right\_click\_count & Right-click events in the window. \\
scroll\_count & Scroll events in the window. \\
click\_count & Left, right, or double-click events in the window. \\
click\_rate & Click events per second. \\
total\_distance & Sum of non-negative mouse movement distances reported in the source CSV. \\
net\_displacement & Euclidean distance between first and last mouse positions in the window. \\
straightness\_ratio & Net displacement divided by total movement distance. \\
mean\_speed & Mean non-negative mouse speed. \\
std\_speed & Standard deviation of non-negative mouse speed. \\
max\_speed & Maximum non-negative mouse speed. \\
mean\_x & Mean mouse x-coordinate. \\
std\_x & Standard deviation of mouse x-coordinate. \\
mean\_y & Mean mouse y-coordinate. \\
std\_y & Standard deviation of mouse y-coordinate. \\
key\_event\_count & Keyboard rows in the window. \\
keypress\_rate & Keyboard rows per second. \\
unique\_key\_count & Number of distinct key labels in the window. \\
wasd\_count & W, A, S, or D key events in the window. \\
wasd\_rate & WASD key events per second. \\
ability\_key\_count & Simple VALORANT ability/equipment key count for q/e/r/f/x/c/z. \\
ability\_rate & Ability/equipment key events per second. \\
modifier\_key\_count & Modifier key events in the window. \\
hold\_mean & Mean key hold duration. \\
hold\_std & Standard deviation of key hold duration. \\
iki\_mean & Mean inter-key interval: next key start minus previous key release. \\
iki\_std & Standard deviation of inter-key intervals. \\
tap\_ratio & Fraction of key holds shorter than 50 ms. \\
action\_intensity & WASD plus ability/equipment key events per second. \\
\end{longtable}

\newpage
\section{Definition of Evaluation Metrics}
\label{app:metrics_definitions}

To evaluate the performance of continuous identification and authentication within the BEACON dataset, four standard biometric and signal detection metrics were utilized. These metrics provide a multi-dimensional view of model reliability, sensitivity, and error distribution.

\begin{itemize}
    \item \textbf{Equal Error Rate (EER):} The EER is the most critical metric for biometric security systems. It represents the specific threshold point where the False Acceptance Rate (FAR) and the False Rejection Rate (FRR) are equal. In the context of the BEACON dataset, a lower EER indicates that the model can effectively distinguish between a legitimate player and an imposter with minimal overlap between their behavioral distributions.
    
    \item \textbf{d-prime ($d'$):} Known as the sensitivity index in Signal Detection Theory, $d'$ measures the statistical distance between the means of the target (genuine user) and noise (imposter) distributions in units of standard deviation. It is calculated as the difference between the Z-scores of the hit rate and the false alarm rate. A higher $d'$ value indicates that the player's behavioral "signal" is significantly stronger than the background "noise" of the general population, regardless of the decision threshold.
    
    \item \textbf{ROC AUC:} The Area Under the Receiver Operating Characteristic Curve provides a single-figure measure of the model's ability to discriminate between classes across all possible thresholds. An AUC of 1.0 represents a perfect classifier, while an AUC of 0.5 suggests performance no better than random chance.
    
    \item \textbf{Accuracy (\%):} In our closed-world identification task, this represents the percentage of temporal windows (10s--60s) where the model correctly identified the specific player out of the 28 available participants.
\end{itemize}

\newpage
\section{Baseline Evaluation Metrics: Detailed Results}
\label{app:detailed_results_table}

Table \ref{tab:results_detailed} provides the complete quantitative results for the 28-class user identification task across all evaluated architectures, temporal windows, and modalities.

\setlength\LTleft{0pt}
\setlength\LTright{0pt}

\small
\begin{longtable}{@{\extracolsep{\fill}} llccccc @{}}
\caption{Multi-modal 28-class Identification Results across Temporal Windows}
\label{tab:results_detailed} \\

\toprule
\textbf{Modality} & \textbf{Model} & \textbf{Window (s)} & \textbf{Acc. (\%)} & \textbf{EER (\%)} & \textbf{d'} & \textbf{AUC} \\
\midrule
\endfirsthead

\multicolumn{7}{c}{{\bfseries \tablename\ \thetable{} -- continued from previous page}} \\
\toprule
\textbf{Modality} & \textbf{Model} & \textbf{Window (s)} & \textbf{Acc. (\%)} & \textbf{EER (\%)} & \textbf{d'} & \textbf{AUC} \\
\midrule
\endhead

\midrule
\multicolumn{7}{r}{{Continued on next page...}} \\
\endfoot

\bottomrule
\endlastfoot

\textbf{Mouse Only} 
 & ARES & 10 & 0.00 & 16.27 & 1.976 & 0.576 \\
 &  & 30 & 0.00 & 28.23 & 1.497 & 0.688 \\
 &  & 45 & 0.00 & 26.50 & 1.643 & 0.397 \\
 &  & 60 & 0.00 & 28.16 & 1.346 & 0.728 \\ \cmidrule{2-7}
 & BAPM & 10 & 46.68 & 5.48 & 2.724 & 0.860 \\
 &  & 30 & 46.12 & 4.71 & 2.780 & 0.889 \\
 &  & 45 & 48.38 & 4.76 & 2.890 & 0.901 \\
 &  & 60 & 54.01 & 4.88 & 2.799 & 0.885 \\ \cmidrule{2-7}
 & NetCLR & 10 & 54.90 & 5.73 & 2.755 & 0.861 \\
 &  & 30 & 54.09 & 5.93 & 2.635 & 0.880 \\
 &  & 45 & 57.43 & 6.14 & 2.714 & 0.877 \\
 &  & 60 & 54.88 & 5.55 & 2.670 & 0.894 \\ \cmidrule{2-7}
 & TCN & 10 & 46.87 & 6.13 & 2.636 & 0.865 \\
 &  & 30 & 44.94 & 8.01 & 2.219 & 0.870 \\
 &  & 45 & 42.42 & 7.64 & 2.232 & 0.859 \\
 &  & 60 & 39.42 & 8.13 & 2.167 & 0.862 \\ \cmidrule{2-7}
 & TMWF & 10 & 53.28 & 6.16 & 2.703 & 0.865 \\
 &  & 30 & 54.63 & 7.61 & 2.423 & 0.862 \\
 &  & 45 & 57.96 & 6.99 & 2.529 & 0.872 \\
 &  & 60 & 56.23 & 6.94 & 2.563 & 0.881 \\ \cmidrule{2-7}
 & Var-CNN & 10 & \textbf{63.16} & \textbf{5.71} & \textbf{2.878} & 0.873 \\
 &  & 30 & 55.01 & 5.95 & 2.697 & 0.868 \\
 &  & 45 & 57.51 & 5.44 & 2.824 & 0.872 \\
 &  & 60 & 50.72 & 5.61 & 2.624 & \textbf{0.901} \\ 
\midrule

\textbf{Keyboard Only} 
 & ARES & 10 & 0.00 & 21.03 & 1.808 & 0.629 \\
 &  & 30 & 0.00 & 28.22 & 1.607 & 0.330 \\
 &  & 45 & 0.00 & 30.62 & 1.355 & 0.648 \\
 &  & 60 & 0.00 & 22.63 & 1.523 & 0.724 \\ \cmidrule{2-7}
 & BAPM & 10 & 16.73 & 11.79 & 1.859 & 0.797 \\
 &  & 30 & 24.52 & 12.61 & 1.687 & 0.847 \\
 &  & 45 & 24.68 & 11.54 & 1.668 & 0.846 \\
 &  & 60 & 22.90 & 11.75 & 1.762 & 0.855 \\ \cmidrule{2-7}
 & NetCLR & 10 & 13.53 & 12.23 & 1.886 & 0.805 \\
 &  & 30 & 24.03 & 13.06 & 1.726 & 0.837 \\
 &  & 45 & 29.21 & 12.80 & 1.732 & 0.851 \\
 &  & 60 & 22.80 & 12.32 & 1.773 & 0.850 \\ \cmidrule{2-7}
 & TCN & 10 & 6.58 & 13.94 & 1.851 & 0.797 \\
 &  & 30 & 20.10 & 14.08 & 1.639 & 0.838 \\
 &  & 45 & 24.75 & 14.12 & 1.645 & 0.857 \\
 &  & 60 & 19.61 & 13.63 & 1.708 & 0.885 \\ \cmidrule{2-7}
 & TMWF & 10 & 13.82 & 12.28 & 1.885 & 0.803 \\
 &  & 30 & 30.01 & 13.23 & 1.793 & 0.840 \\
 &  & 45 & \textbf{36.23} & \textbf{11.54} & \textbf{1.886} & \textbf{0.885} \\
 &  & 60 & 32.66 & 13.06 & 1.801 & 0.846 \\ \cmidrule{2-7}
 & Var-CNN & 10 & 15.56 & 11.79 & 1.882 & 0.801 \\
 &  & 30 & 26.24 & 12.52 & 1.747 & 0.841 \\
 &  & 45 & 27.77 & 12.09 & 1.776 & 0.855 \\
 &  & 60 & 24.35 & 13.00 & 1.758 & 0.848 \\ 
 \midrule

\makecell{\textbf{Combined Mouse} \\ \textbf{and Keyboard}} 
 & ARES & 10 & 0.00 & 18.52 & 1.863 & 0.566 \\
 &  & 30 & 0.00 & 23.80 & 1.516 & 0.727 \\
 &  & 45 & 0.00 & 28.28 & 1.312 & 0.746 \\
 &  & 60 & 0.00 & 23.34 & 1.592 & 0.232 \\ \cmidrule{2-7}
 & BAPM & 10 & 47.18 & 4.75 & 2.922 & 0.875 \\
 &  & 30 & 54.47 & 4.82 & 3.039 & 0.882 \\
 &  & 45 & 55.32 & 4.50 & 3.116 & 0.902 \\
 &  & 60 & 61.93 & 4.31 & 3.192 & 0.902 \\ \cmidrule{2-7}
 & NetCLR & 10 & 62.16 & 5.48 & 2.876 & 0.874 \\
 &  & 30 & 66.54 & 6.21 & 2.821 & 0.880 \\
 &  & 45 & 66.11 & 5.05 & 3.042 & 0.884 \\
 &  & 60 & 69.08 & 5.19 & 3.016 & 0.911 \\ \cmidrule{2-7}
 & TCN & 10 & 55.54 & 6.10 & 2.698 & 0.872 \\
 &  & 30 & 46.44 & 7.09 & 2.404 & 0.864 \\
 &  & 45 & 45.13 & 7.77 & 2.299 & 0.878 \\
 &  & 60 & 41.84 & 7.26 & 2.327 & 0.887 \\ \cmidrule{2-7}
 & TMWF & 10 & 62.33 & 7.04 & 2.703 & 0.856 \\
 &  & 30 & 58.03 & 6.67 & 2.604 & 0.867 \\
 &  & 45 & 66.11 & 6.13 & 2.851 & 0.873 \\
 &  & 60 & 69.66 & 6.77 & 2.737 & 0.889 \\ \cmidrule{2-7}
 & Var-CNN & 10 & 59.52 & 5.16 & 2.936 & 0.877 \\
 &  & 30 & 53.12 & 5.46 & 2.782 & 0.896 \\
 &  & 45 & 68.53 & 4.68 & 3.072 & 0.910 \\
 &  & 60 & \textbf{70.82} & \textbf{4.31} & \textbf{3.192} & \textbf{0.911} \\ 

\end{longtable}

\newpage
\section{Baseline Evaluation Performance Curves}
\label{app:evaluation_graphs}

This appendix provides a granular visual and analytical breakdown of the performance characteristics for the six evaluated Website Fingerprinting (WF) architectures. The following sections visualize the trade-offs between temporal observation windows (10s, 30s, 45s, and 60s) and the resulting biometric separability.

\subsection{Unimodal Mouse Dynamics}
Mouse telemetry provides a continuous high-fidelity stream of sensorimotor data, resulting in significantly higher identification benchmarks.

\begin{figure}[!ht]
    \centering
    \begin{subfigure}[b]{0.48\textwidth}
        \centering
        \includegraphics[page=1, width=\textwidth]{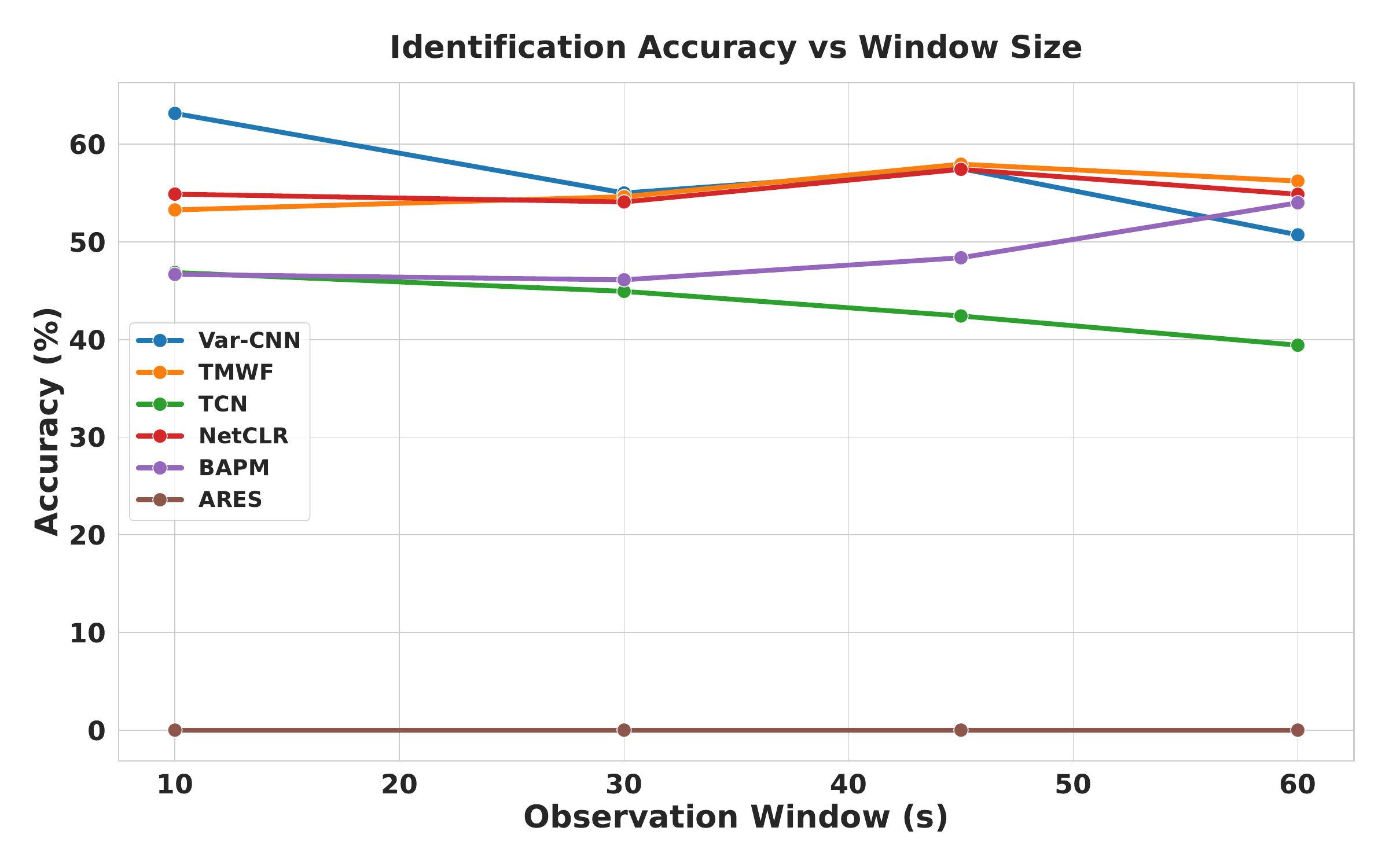}
        \caption{Accuracy vs. Window Size}
    \end{subfigure}
    \hfill
    \begin{subfigure}[b]{0.48\textwidth}
        \centering
        \includegraphics[page=2, width=\textwidth]{mouse_res.pdf}
        \caption{Equal Error Rate (EER) Trends}
    \end{subfigure}
    \vspace{1em}
    \begin{subfigure}[b]{0.48\textwidth}
        \centering
        \includegraphics[page=3, width=\textwidth]{mouse_res.pdf}
        \caption{Biometric Separation ($d'$)}
    \end{subfigure}
    \hfill
    \begin{subfigure}[b]{0.48\textwidth}
        \centering
        \includegraphics[page=4, width=\textwidth]{mouse_res.pdf}
        \caption{Performance Heatmap}
    \end{subfigure}
    \caption{Detailed Performance Metrics for Unimodal Mouse Dynamics.}
    \label{fig:app_mouse_detailed}
\end{figure}

\textbf{Analytical Summary:} Figure \ref{fig:app_mouse_detailed} highlights the rapid convergence of Var-CNN and NetCLR. Even at a 10s resolution, mouse dynamics yield identification accuracies exceeding 50\%. This is due to the extreme data density of the cursor polling rate ($>$500Hz), which provides the models with sufficient samples to model aim-path curvature and acceleration profiles traits that are highly individualized in high-skill FPS environments.

\newpage

\subsection{Unimodal Keyboard Dynamics}
The keyboard modality represents the most challenging identification task due to the discrete and sparse nature of the input stream. 

\begin{figure}[!ht]
    \centering
    \begin{subfigure}[b]{0.48\textwidth}
        \centering
        \includegraphics[page=1, width=\textwidth]{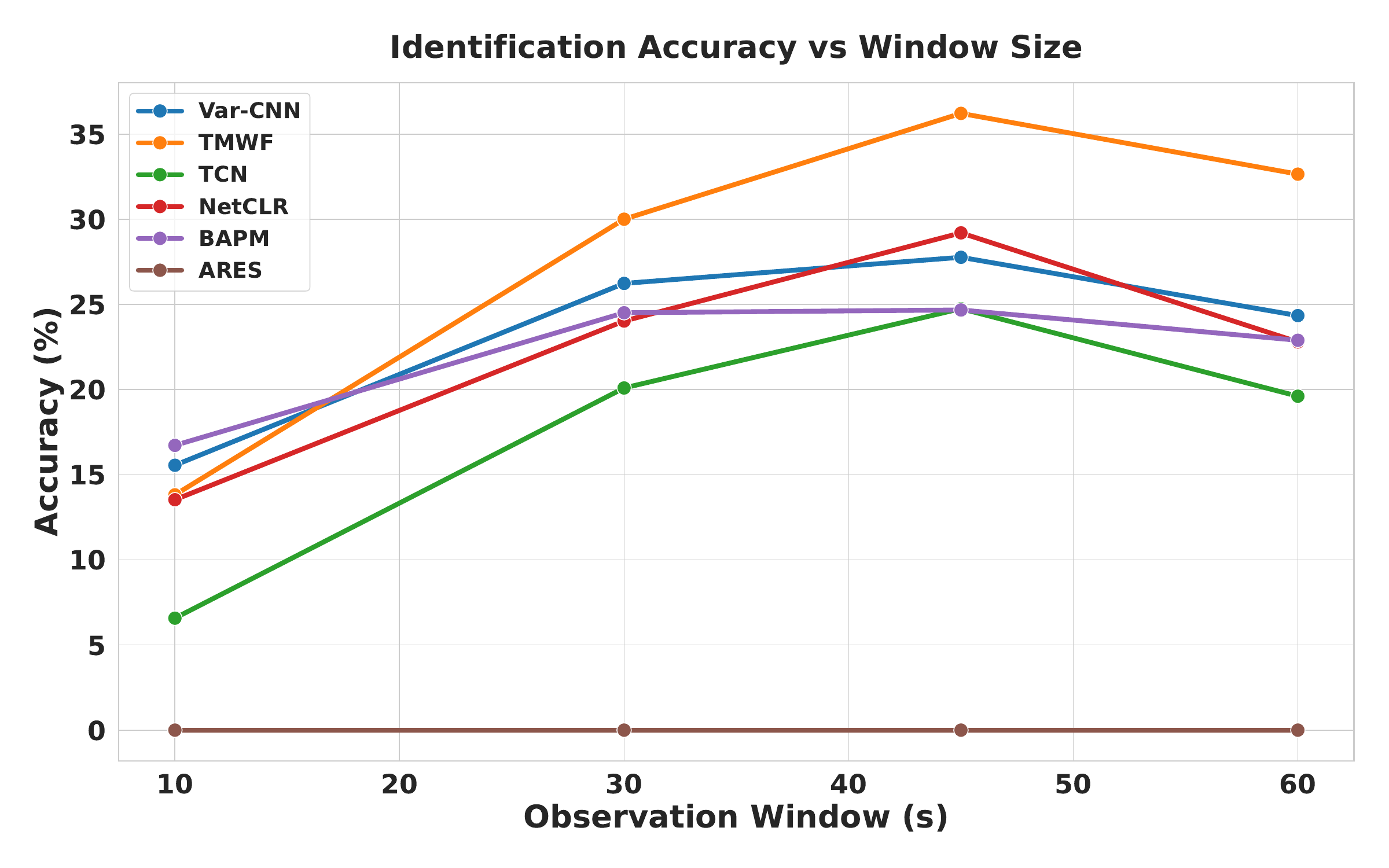}
        \caption{Accuracy vs. Window Size}
    \end{subfigure}
    \hfill
    \begin{subfigure}[b]{0.48\textwidth}
        \centering
        \includegraphics[page=2, width=\textwidth]{keyboard_res.pdf}
        \caption{Equal Error Rate (EER) Trends}
    \end{subfigure}
    \vspace{1em}
    \begin{subfigure}[b]{0.48\textwidth}
        \centering
        \includegraphics[page=3, width=\textwidth]{keyboard_res.pdf}
        \caption{Biometric Separation ($d'$)}
    \end{subfigure}
    \hfill
    \begin{subfigure}[b]{0.48\textwidth}
        \centering
        \includegraphics[page=4, width=\textwidth]{keyboard_res.pdf}
        \caption{Performance Heatmap}
    \end{subfigure}
    \caption{Detailed Performance Metrics for Unimodal Keyboard Dynamics.}
    \label{fig:app_keyboard_detailed}
\end{figure}

\textbf{Analytical Summary:} As observed in Figure \ref{fig:app_keyboard_detailed}, identification accuracy for keyboard dynamics remains relatively low ($<$40\%) compared to other modalities. This is attributed to the "high-latency" nature of tactical binds; in shorter windows, a participant may only trigger a limited number of unique keys, leading to feature sparsity. However, the $d'$ score improvement at 60s suggests that longer observation windows successfully capture idiosyncratic typing rhythms (dwell times and IKIs) that are otherwise obscured in shorter intervals.

\newpage

\subsection{Multi-modal Early Fusion (Combined)}
The fusion of mouse and keyboard telemetry represents the upper bound of the BEACON dataset's biometric separability.

\begin{figure}[!ht]
    \centering
    \begin{subfigure}[b]{0.48\textwidth}
        \centering
        \includegraphics[page=1, width=\textwidth]{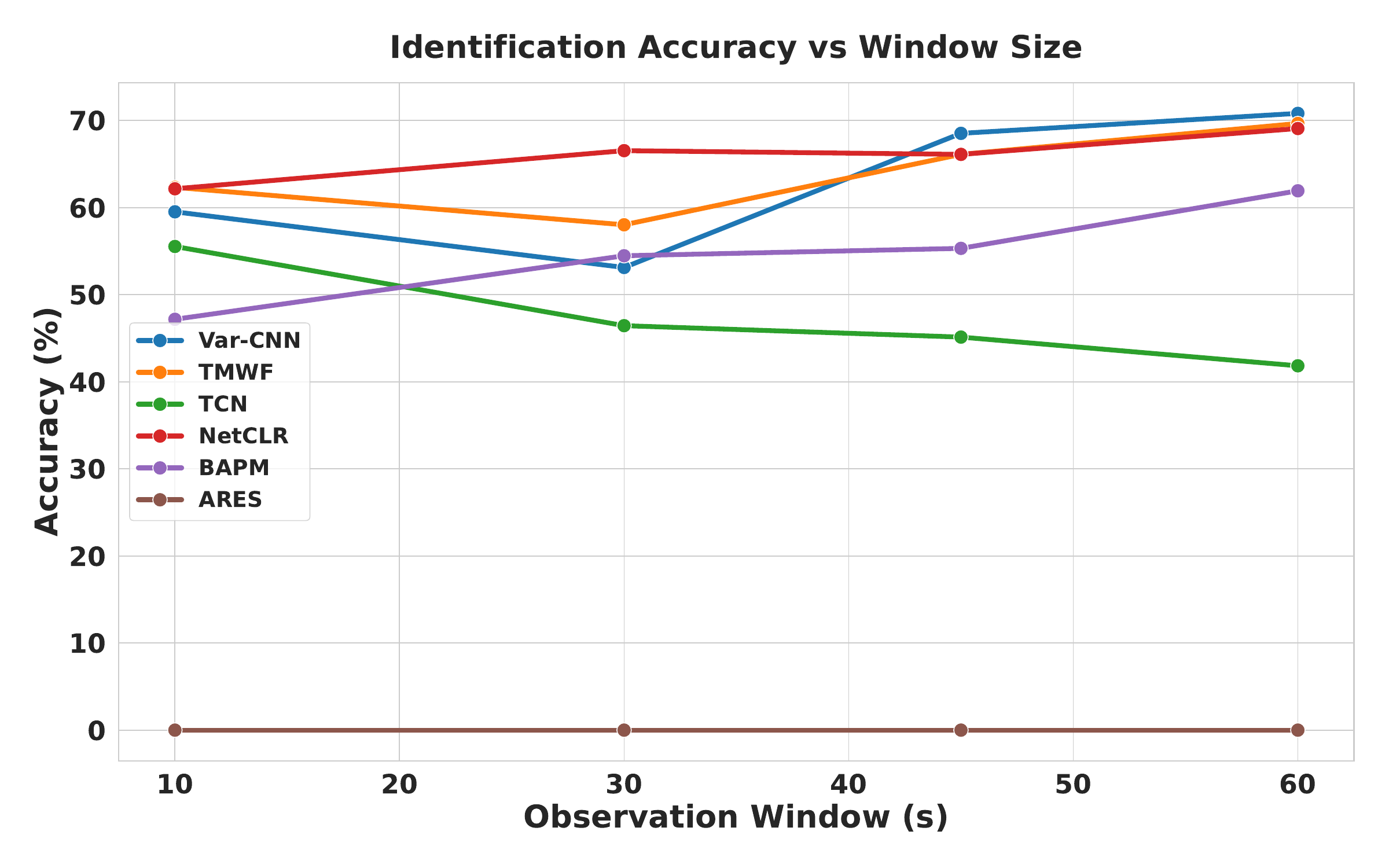}
        \caption{Accuracy vs. Window Size}
    \end{subfigure}
    \hfill
    \begin{subfigure}[b]{0.48\textwidth}
        \centering
        \includegraphics[page=2, width=\textwidth]{combined_res.pdf}
        \caption{Equal Error Rate (EER) Trends}
    \end{subfigure}
    \vspace{1em}
    \begin{subfigure}[b]{0.48\textwidth}
        \centering
        \includegraphics[page=3, width=\textwidth]{combined_res.pdf}
        \caption{Biometric Separation ($d'$)}
    \end{subfigure}
    \hfill
    \begin{subfigure}[b]{0.48\textwidth}
        \centering
        \includegraphics[page=4, width=\textwidth]{combined_res.pdf}
        \caption{Performance Heatmap}
    \end{subfigure}
    \caption{Detailed Performance Metrics for Multi-modal Early Fusion (Combined Profile).}
    \label{fig:app_fusion_detailed}
\end{figure}

\textbf{Analytical Summary:} The combined results in Figure \ref{fig:app_fusion_detailed} demonstrate a synergistic effect. While mouse data provides the core distinctiveness, the addition of keyboard tactical bind ratios acts as a "behavioral stabilizer," reducing the EER to a global minimum of 4.31\% at 60s. The failure of the ARES model across all pages indicates a specific incompatibility between its automated representation learning layers and the engineered feature vectors, highlighting that architectures optimized for packet-burst patterns in Tor traffic require significant re-tuning for human sensorimotor dynamics.

\newpage

\section{Datasheet for the BEACON Dataset}
\label{app:datasheet}

This appendix provides a structured datasheet for BEACON. It consolidates motivation, composition, collection process, preprocessing, intended uses, distribution, and maintenance information that supports informed downstream use of the dataset.

\subsection{Motivation}

BEACON was created to support research into passive, continuous behavioral biometric authentication in competitive first-person-shooter (FPS) gaming. Existing behavioral-biometric corpora are dominated by desktop productivity tasks (typing, web browsing) or single-modality mobile-touch studies, and they do not capture the high-frequency, multi-modal telemetry characteristic of competitive gameplay. BEACON addresses this gap by releasing synchronised mouse dynamics, keystroke dynamics, network telemetry, screen recordings, and hardware context from real \textit{Valorant} sessions. The dataset was assembled by an academic research team as part of a supervised undergraduate capstone project, under institutional ethical oversight, with no external funding.

\subsection{Composition}

Each instance is a single \textit{Valorant} gameplay session contributed by one participant. A canonical session contains up to five synchronised artefacts: a screen recording (\texttt{.mp4}), a network capture (\texttt{.pcap}), a mouse event log (\texttt{.csv}), a keystroke log (\texttt{.csv}), and a hardware context snapshot (\texttt{.json}). The release contains 79 sessions from 28 distinct participants, with sessions per participant ranging from 1 to 19 (mean $\approx 2.8$). The participant pool is a convenience sample drawn from a single university and is not intended to be statistically representative of the broader \textit{Valorant} player population.

Each session is labelled with a participant identity in the form of an opaque pseudonymous identifier (P001--P028); raw participant names are mapped to these IDs through a private salt-keyed table held only by the curators. No match-level outcome labels (win/loss, kills/deaths) are included.

The deviations between session count (79) and per-modality file counts are documented in the footnote of Table~\ref{tab:modality_inventory}: 7 sessions lack screen recordings (4 early pilot sessions with capture intentionally disabled, 3 capture-utility failures), 2 sessions yielded split-part videos, and one session folder bundles two distinct recording timestamps for a single participant, producing two files per input modality and one missing hardware JSON.

The following information is \emph{not} captured per session and constitutes a known gap relative to the Gebru et al.\ recommended fields: per-session \textit{Valorant} patch version, in-game rank, mouse DPI/polling rate, in-game sensitivity, structured age, structured gender, handedness, and geographic location beyond the host institution. Sessions span 16~April~2025 to 20~October~2025, encompassing \textit{Valorant} patches v10.07 through v11.08 across Season 2025 Acts 2--6; the active patch was not logged per session. Self-reported metadata at recruitment indicated a cohort of university-aged students (estimated 18--25), 27 male and 1 female, with skill tiers described as ranging from unranked to Diamond.

Known noise sources and minor irregularities include: a small number of non-monotonic or duplicate mouse timestamps, documented per-session in the released feature summary tables; preserved \texttt{BackupKeybinds.json} files inside auxiliary \textit{Valorant} configuration directories on the shared lab machine, which are excluded from the canonical modality inventory; and two superseded participant folders that are present on disk but excluded from all analysis scripts via an explicit ignore list.

The raw hardware JSONs contain machine-level identifiers (hostname, IP, MAC) that are scrubbed in the public anonymised release; the unredacted raw data are not redistributed. Screen recordings may incidentally include on-screen content beyond the game (e.g., desktop notifications, taskbar items), as discussed in the Ethics Statement.

\subsection{Collection Process}

A custom Python logging agent (\S\ref{sec:logger_arch}) was deployed on the recording workstation. It runs four concurrent monitors: a global keyboard hook (\texttt{pynput.keyboard}), a global mouse hook (\texttt{pynput.mouse}), a packet-capture sniffer (\texttt{scapy.sniff} backed by libpcap/Npcap), and a screen recorder (\texttt{FFmpeg} subprocess at 25~fps with \texttt{libx264 -preset ultrafast -pix\_fmt yuv420p}). A hardware-context script (\texttt{wmi}/\texttt{psutil} queries) fires once at session start and writes the corresponding JSON. All four streams are stamped with the host's POSIX time (Python \texttt{time.time()}), providing a single shared clock that allows post-hoc temporal alignment without an explicit synchronisation marker.

Participants were invited to a controlled laboratory environment and played \textit{Valorant} in unconstrained sessions; no specific game mode, map, or task was mandated. Sessions were terminated by the participant or supervising researcher. Recruitment was conducted through the host institution; all participants provided informed written consent prior to data capture, and a per-session consent artefact (\texttt{consent\_granted\_[timestamp].txt}) is retained with the raw data.

\subsection{Hardware Configurations}

Consistent with the hybrid collection approach described in \S\ref{sec:logger_arch}, the majority of sessions were recorded on two standardised laboratory desktop configurations, while a minority of sessions were contributed from participants' home machines to add real-world hardware variety:

\begin{itemize}
    \item \textbf{Setup 1 (lab, majority of sessions).} Intel Core~i7 (12th-generation Alder Lake-class), NVIDIA GeForce GTX~1070 discrete GPU, standard chiclet keyboard, Razer DeathAdder mouse, 1920$\times$1080 monitor, cloth mousepad.
    \item \textbf{Setup 2 (lab).} Intel Core~i7 with integrated graphics (no discrete GPU), standard chiclet keyboard, Redgear Dragonwar Ele-G9 mouse, 1920$\times$1080 monitor, cloth mousepad.
    \item \textbf{Home contributions.} A small number of sessions were captured on participants' personal desktops; the released hardware JSONs accordingly include additional CPU fingerprints (notably Zen~4 Ryzen and 13th/14th-generation Intel Core families) that originate from these home sessions rather than the standardised lab benches.
\end{itemize}

Audio peripherals (in-ear monitors, headphones, or earphones) were not standardised; participants supplied their own. Mouse DPI, polling rate, and per-game sensitivity settings were not centrally captured, although physical device VIDs/PIDs are recorded inside the per-session hardware JSON. The published hardware JSONs additionally report observed CPU family/model strings, RAM, OS version, display configuration, and detected I/O devices.

\subsection{Preprocessing, Cleaning, and Labelling}

The released artefacts include per-session modality files (after the anonymisation pipeline described in \S\ref{subsec:datasheet_distribution}) together with derived feature summary tables used for the baselines reported in this paper. Filtering bounds applied during downstream feature extraction are conservative and explicitly documented: mouse-speed values outside $[0, 12{,}000]$~px/s and keystroke dwell durations outside $[0, 5{,}000]$~ms are excluded from feature computation; whole-session CSVs are used for the published analysis (per-match split CSVs, where present, are excluded but available); and PCAP feature extraction sub-samples up to $5{\times}10^{5}$ packets per file for tractability during baseline computation. This sub-sampling is a feature-extraction step only and does not affect the packet counts of the anonymised PCAPs distributed in the release. All preprocessing scripts that reproduce the paper tables and figures from raw artefacts are included in the released code repository.

\subsection{Uses}

BEACON has been used in this paper to benchmark behavioral biometric identification using mouse, keyboard, and early-fusion combined modalities under several deep architectures (\S\ref{sec:results}). Beyond this, the dataset is intended to support research on continuous authentication, longitudinal behavioral stability and drift, multi-modal biometric fusion, mimicry- and bot-detection, FPS performance modelling, and gameplay-aware network traffic analysis.

The cohort composition imposes important constraints on downstream use. The shared-laboratory setup, single-institution recruitment, narrow age range, and skewed gender distribution mean that in-lab patterns are unlikely to match home-environment behaviour, and that any conclusions drawn must be qualified accordingly. Patch-induced gameplay drift across the six-month collection window is also unmodelled. Concretely, the dataset should \emph{not} be used for real-world re-identification of participants from the unredacted raw data, nor for training deployable anti-cheat or surveillance systems without further validation on broader populations.

\subsection{Distribution}
\label{subsec:datasheet_distribution}

An anonymised release of BEACON covering all 28 participants (P001--P028) is hosted on Hugging Face Datasets. Raw data containing machine-level identifiers is retained internally by the curators and is not redistributed. This dataset is licensed under a Creative Commons Attribution-NonCommercial 4.0 International License (CC BY-NC 4.0).

The anonymisation pipeline, whose source is released alongside the dataset, applies modality-specific transformations:

\begin{itemize}
    \item \textbf{Identity mapping.} Raw participant names and session timestamps are replaced with opaque IDs (\texttt{P001}--\texttt{P028}, \texttt{S001}, \texttt{S002}, \ldots) via a private salt-keyed mapping that is never published. Filenames and on-disk paths are rewritten accordingly.
    \item \textbf{Hardware JSONs.} Hostnames, IPv4/IPv6, and MAC addresses are removed and any regex-detectable identifier strings inside nested device fields are replaced with redaction tokens. CPU family/model strings, RAM, OS version, monitor geometry, and device-class VIDs/PIDs are retained.
    \item \textbf{Network captures.} PCAPs are not redistributed in raw form. IPv4 source/destination addresses are deterministically remapped via a salt-keyed hash into a private \texttt{10.0.0.0/8} range, IPv6 addresses are replaced with fixed unique-local placeholders, link-layer MAC addresses are remapped, and IP checksums are recomputed. Local-network discovery traffic on UDP ports known to leak hostnames in plaintext (mDNS, NetBIOS, SSDP) is dropped entirely. The dropped packet counts are logged per session.
    \item \textbf{Screen recordings.} Videos are released with audio removed, container metadata stripped, and on-screen regions known to leak identifying content masked. The current pipeline applies coarse top/bottom region masking via an \texttt{FFmpeg} drawbox filter; a region-targeted pass that masks identifying in-game UI elements and any host-OS surfaces visible during alt-tabs is under active development and will be applied to all videos prior to the public release. Sessions for which redaction cannot be reliably applied will be omitted from the public video subset rather than released unredacted.
    \item \textbf{Consent artefacts.} Raw \texttt{consent\_granted\_[timestamp].txt} files captured at session start may contain direct identifiers and are therefore not part of the public release; they are retained privately to support audit and withdrawal requests.
\end{itemize}

Network traffic captures contain only \textit{Valorant}'s encrypted protocol data, and screen recordings depict the game's normal visual output. Researchers should independently verify compliance with Riot Games' terms of service for their specific use case.

\subsection{Maintenance}

The dataset is hosted and maintained by the curating research team. Future revisions may add participant sessions and supplementary metadata; versioned snapshots will be preserved on the Hugging Face hub. The private participant mapping table supports targeted withdrawal: an enrolled participant may request removal of all of their sessions without disrupting the remainder of the corpus. Contact information for data-handling requests will be provided alongside the dataset card at release time.

\newpage
\section*{NeurIPS Paper Checklist}

\begin{enumerate}

\item {\bf Claims}
    \item[] Question: Do the main claims made in the abstract and introduction accurately reflect the paper's contributions and scope?
    \item[] Answer: \answerYes{}
    \item[] Justification: The abstract and introduction clearly state that BEACON is a dataset
    contribution, describe its scale and modalities, and accurately reflect the baseline
    evaluation scope presented in Section~4. No claims are made beyond what is demonstrated.

\item {\bf Limitations}
    \item[] Question: Does the paper discuss the limitations of the work performed by the authors?
    \item[] Answer: \answerYes{}
    \item[] Justification: A dedicated Limitations section (Section~6) discusses cohort size,
    single-game/platform scope, exclusion of PCAP and video modalities from baselines, and
    the absence of longitudinal data.

\item {\bf Theory assumptions and proofs}
    \item[] Question: For each theoretical result, does the paper provide the full set of assumptions and a complete (and correct) proof?
    \item[] Answer: \answerNA{}
    \item[] Justification: This paper introduces a dataset and empirical baselines; it contains
    no theoretical results, theorems, or formal proofs.

\item {\bf Experimental result reproducibility}
    \item[] Question: Does the paper fully disclose all the information needed to reproduce the main experimental results of the paper to the extent that it affects the main claims and/or conclusions of the paper (regardless of whether the code and data are provided or not)?
    \item[] Answer: \answerYes{}
    \item[] Justification: Section~4 specifies the model architectures, optimizer, learning rate,
    batch size, sequence length, train/val/test splits, temporal window sizes, and confidence
    threshold. The dataset is publicly released on HuggingFace and the logger codebase on
    Zenodo.

\item {\bf Open access to data and code}
    \item[] Question: Does the paper provide open access to the data and code, with sufficient instructions to faithfully reproduce the main experimental results, as described in supplemental material?
    \item[] Answer: \answerYes{}
    \item[] Justification: The BEACON dataset is publicly available at
    \url{https://huggingface.co/datasets/beacon-gui/BEACON-Dataset} and the data collection
    logger is released at \url{https://zenodo.org/records/20062628}.

\item {\bf Experimental setting/details}
    \item[] Question: Does the paper specify all the training and test details (e.g., data splits, hyperparameters, how they were chosen, type of optimizer) necessary to understand the results?
    \item[] Answer: \answerYes{}
    \item[] Justification: Section~4 provides the full experimental setup including Adam optimizer
    (lr=$10^{-3}$), CrossEntropyLoss, 30 epochs, batch size 32, 80/10/20 chronological
    data splits, sequence length 1024, and window sizes of 10s, 30s, 45s, and 60s.

\item {\bf Experiment statistical significance}
    \item[] Question: Does the paper report error bars suitably and correctly defined or other appropriate information about the statistical significance of the experiments?
    \item[] Answer: \answerNo{}
    \item[] Justification: Error bars are not reported in Table~4 as each architecture was
    evaluated in a single deterministic run under fixed chronological splits. Reporting
    variance across multiple random seeds was computationally prohibitive given six
    architectures across three modalities and four temporal resolutions.

\item {\bf Experiments compute resources}
    \item[] Question: For each experiment, does the paper provide sufficient information on the computer resources (type of compute workers, memory, time of execution) needed to reproduce the experiments?
    \item[] Answer: \answerYes{}
    \item[] Justification: Section~4 states that all experiments were conducted on an NVIDIA
    H100 80GB GPU provided by the Thapar School of Advanced AI and Data Science.
    Individual run times are not reported but the hyperparameter configuration is fully
    specified to allow estimation.

\item {\bf Code of ethics}
    \item[] Question: Does the research conducted in the paper conform, in every respect, with the NeurIPS Code of Ethics?
    \item[] Answer: \answerYes{}
    \item[] Justification: All participants provided informed written consent. Data collection
    was conducted in accordance with the ethical guidelines of Thapar Institute of
    Engineering \& Technology. The released dataset is fully anonymized with no personally
    identifiable information retained.

\item {\bf Broader impacts}
    \item[] Question: Does the paper discuss both potential positive societal impacts and negative societal impacts of the work performed?
    \item[] Answer: \answerYes{}
    \item[] Justification: The Future Scope section (Section~8) discusses positive applications
    including continuous authentication, fatigue detection, and skill assessment. The
    Limitations section acknowledges potential dual-use risks such as behavioral surveillance;
    the dataset's anonymization mitigates direct privacy harms.

\item {\bf Safeguards}
    \item[] Question: Does the paper describe safeguards that have been put in place for responsible release of data or models that have a high risk for misuse (e.g., pre-trained language models, image generators, or scraped datasets)?
    \item[] Answer: \answerYes{}
    \item[] Justification: The dataset is released in anonymized form with participant
    pseudonyms (P001--P028) and no personally identifiable information. Data collection
    involved explicit informed consent from all participants, and the release is governed
    by the HuggingFace dataset hosting terms of service.

\item {\bf Licenses for existing assets}
    \item[] Question: Are the creators or original owners of assets (e.g., code, data, models) used in the paper properly credited and are the license and terms of use explicitly mentioned and properly respected?
    \item[] Answer: \answerYes{}
    \item[] Justification: All six baseline architectures (ARES, BAPM, NetCLR, TCN, TMWF,
    Var-CNN) are properly cited in Section~4. The Valorant game platform is cited and
    credited to Riot Games \cite{riot_valorant_2020}. All referenced datasets in Table~1
    are cited with their original publications.

\item {\bf New assets}
    \item[] Question: Are new assets introduced in the paper well documented and is the documentation provided alongside the assets?
    \item[] Answer: \answerYes{}
    \item[] Justification: The BEACON dataset is documented via modality inventory (Table~2),
    feature engineering dictionary (Appendix~C), and EDA (Section~3). The BEACON logger
    codebase is released on Zenodo with architecture description in Section~2.1 and
    Appendix~A.

\item {\bf Crowdsourcing and research with human subjects}
    \item[] Question: For crowdsourcing experiments and research with human subjects, does the paper include the full text of instructions given to participants and screenshots, if applicable, as well as details about compensation (if any)?
    \item[] Answer: \answerNo{}
    \item[] Justification: The paper confirms informed written consent was obtained from all
    28 participants (Section~2). Full participant instruction scripts are not included in
    the paper but were administered prior to each session. Participation was voluntary;
    no monetary compensation was provided.

\item {\bf Institutional review board (IRB) approvals or equivalent for research with human subjects}
    \item[] Question: Does the paper describe potential risks incurred by study participants, whether such risks were disclosed to the subjects, and whether Institutional Review Board (IRB) approvals (or an equivalent approval/review based on the requirements of your country or institution) were obtained?
    \item[] Answer: \answerYes{}
    \item[] Justification: Section~2 states that the study was conducted in accordance with
    the ethical guidelines of Thapar Institute of Engineering \& Technology, in compliance
    with institutional policies for human-subjects research. Participants were informed of
    data collection scope prior to consent.

\item {\bf Declaration of LLM usage}
    \item[] Question: Does the paper describe the usage of LLMs if it is an important,
    original, or non-standard component of the core methods in this research? Note that
    if the LLM is used only for writing, editing, or formatting purposes and does
    \emph{not} impact the core methodology, scientific rigor, or originality of the
    research, declaration is not required.
    \item[] Answer: \answerNA{}
    \item[] Justification: LLMs were used only for grammar and writing assistance during
    manuscript preparation and do not form any part of the core methodology.

\end{enumerate}

\end{document}